\documentclass[journal, letterpaper, twoside, 10pt]{IEEEtran}
\IEEEoverridecommandlockouts
\usepackage[bookmarks=false]{hyperref}    
\usepackage[noadjust]{cite}
\usepackage{amsmath,amssymb,amsfonts}
\usepackage{algorithm}
\usepackage{algorithmic}
\usepackage{graphicx}
\usepackage{textcomp}
\usepackage{xcolor}
\def\BibTeX{{\rm B\kern-.05em{\sc i\kern-.025em b}\kern-.08em
    T\kern-.1667em\lower.7ex\hbox{E}\kern-.125emX}}
    
\usepackage{rotating}
\usepackage{adjustbox}
\usepackage{pifont}
\usepackage{framed}
\usepackage{soul}
\usepackage{booktabs}  
\usepackage{csquotes}
\usepackage{multirow}
\usepackage{array}
\usepackage{balance}
\usepackage{wasysym}
\newcolumntype{L}[1]{>{\raggedright\let\newline\\\arraybackslash\hspace{0pt}}m{#1}}
\newcolumntype{C}[1]{>{\centering\let\newline\\\arraybackslash\hspace{0pt}}m{#1}}
\newcolumntype{R}[1]{>{\raggedleft\let\newline\\\arraybackslash\hspace{0pt}}m{#1}}
\usepackage{url}
\MakeOuterQuote{"}

\newcommand{\aftertable}{\vspace{-2em}}
\newcommand{\afterfig}{\vspace{-10pt}}

\usepackage{booktabs} 
\usepackage{pifont}

\newcommand\OK {\textcolor{red}{\ding{55}} }
\newcommand\NO { \textcolor{black}{\ding{51}} } 
\usepackage{enumitem}
\usepackage{subfigure}
\begin{document}
\bstctlcite{IEEEexample:BSTcontrol}

\title{Benchmarking at the Frontier of Hardware Security: Lessons from Logic Locking}

\author{
\IEEEauthorblockN{
\parbox{\linewidth}{\centering
{\bf Coordinators:} Benjamin~Tan\IEEEauthorrefmark{1},
Ramesh~Karri\IEEEauthorrefmark{1}, \\
\textcolor{blue}{{\bf Blue teams:}
Nimisha~Limaye\IEEEauthorrefmark{1},
Abhrajit~Sengupta\IEEEauthorrefmark{1},
Ozgur~Sinanoglu\IEEEauthorrefmark{2} (combinational locking), \\
Md~Moshiur~Rahman\IEEEauthorrefmark{3},
Swarup~Bhunia\IEEEauthorrefmark{3} (sequential locking)}, \\
\textcolor{red}{
{\bf Red teams:}
Danielle~Duvalsaint\IEEEauthorrefmark{4},
R.D.~(Shawn)~Blanton\IEEEauthorrefmark{4},
Amin~Rezaei\IEEEauthorrefmark{5},
Yuanqi~Shen\IEEEauthorrefmark{5},
Hai~Zhou\IEEEauthorrefmark{5},
 Leon~Li\IEEEauthorrefmark{6},
Alex~Orailoglu\IEEEauthorrefmark{6},
Zhaokun~Han\IEEEauthorrefmark{7},
Austin~Benedetti\IEEEauthorrefmark{7},
Luciano~Brignone\IEEEauthorrefmark{7},
Muhammad~Yasin\IEEEauthorrefmark{7},
Jeyavijayan~Rajendran\IEEEauthorrefmark{7},
Michael~Zuzak\IEEEauthorrefmark{8},
Ankur~Srivastava\IEEEauthorrefmark{8},
Ujjwal~Guin\IEEEauthorrefmark{9},
Chandan~Karfa\IEEEauthorrefmark{10},
Kanad~Basu\IEEEauthorrefmark{11}}  \\
{\bf Judges, Evaluators, Community-at-large}: Vivek~V.~Menon\IEEEauthorrefmark{12},
Matthew~French\IEEEauthorrefmark{12},
Peilin~Song\IEEEauthorrefmark{13},
Franco~Stellari\IEEEauthorrefmark{13},
Gi-Joon~Nam\IEEEauthorrefmark{13},
Peter~Gadfort\IEEEauthorrefmark{14},
Alric~Althoff\IEEEauthorrefmark{15},
Joseph~Tostenrude\IEEEauthorrefmark{16},
Saverio~Fazzari\IEEEauthorrefmark{17},
Eric~Breckenfeld\IEEEauthorrefmark{17},
Ken~Plaks\IEEEauthorrefmark{17} \\
}
}
\vspace{10pt}
\IEEEauthorblockA{%
\parbox{\linewidth}{\centering
\IEEEauthorrefmark{1}New~York~University, 
\IEEEauthorrefmark{2}New~York~University~Abu~Dhabi, 
\IEEEauthorrefmark{3}University~of~Florida, 
\IEEEauthorrefmark{4}Carnegie~Mellon~University, 
\IEEEauthorrefmark{5}Northwestern~University, 
\IEEEauthorrefmark{6}University~of~California~San~Diego, 
\IEEEauthorrefmark{7}Texas~A\&M~University, 
\IEEEauthorrefmark{8}University~of~Maryland~College~Park, 
\IEEEauthorrefmark{9}Auburn~University, 
\IEEEauthorrefmark{10}Indian~Institute~of~Technology~Guwahati, 
\IEEEauthorrefmark{11}University~of~Texas~at~Dallas, 
\IEEEauthorrefmark{12}Information~Sciences~Institute~--~University~of~Southern~California, 
\IEEEauthorrefmark{13}IBM~T.~J.~Watson~Research~Center, 
\IEEEauthorrefmark{14}CCDC~Army~Research~Laboratory, 
\IEEEauthorrefmark{15}Tortuga~Logic, 
\IEEEauthorrefmark{16}Boeing, 
 \IEEEauthorrefmark{17}DARPA
}
}\vspace{-20pt}
}

\maketitle
 
\begin{abstract}
Integrated circuits (ICs) are the foundation of all computing systems. 
They comprise high-value hardware intellectual property (IP) that are at risk of piracy, reverse-engineering, and modifications while making their way through the geographically-distributed IC supply chain. On the frontier of hardware security are various design-for-trust techniques that claim to protect designs from untrusted entities across the design flow. Logic locking is one technique that promises protection from the gamut of threats in IC manufacturing. In this work, we perform a critical review of logic locking techniques in the literature, and expose several shortcomings. Taking inspiration from other cybersecurity competitions, we devise a community-led benchmarking exercise to address the evaluation deficiencies. In reflecting on this process, we shed new light on deficiencies in evaluation of logic locking and reveal important future directions. The lessons learned can guide future endeavors in other areas of hardware security. 
\end{abstract}

\section{Introduction}
\label{sec:intro}

\subsection{Hardware Security: Mitigating Threats to the Supply Chain}

    Integrated circuits (ICs) are the foundation of all computing systems, and the need for security spans the entire system life-cycle, from conception, through hardware design and manufacture, and across the system's operating lifetime. 
    Hardware Intellectual Property (IP) protection is crucial, especially as the market for domain-specific computer hardware becomes more lucrative~\cite{golden-era}. 
    Given the high-value effort involved in hardware design, we want to prevent IP theft, counterfeiting, malicious modifications, and reverse-engineering. 
    Effective security techniques for use in the IC supply chain are of the utmost importance for mitigating such threats. 
    
     \begin{table}[!bt]
    \caption{%
    Comparison of selected Design-for-Trust techniques~\cite{Yasin2020-book}. 
    \NO means a technique can prevent piracy by the untrusted entity.
    \label{tab:intro-table}}
    \resizebox{\columnwidth}{!}{
    \begin{tabular}{@{}lC{1cm}C{1cm}C{1cm}C{1cm}@{}}
    \toprule
    & \multicolumn{4}{c}{Untrusted Entity} \\ \cmidrule(l){2-5} 
    Design-for-Trust Technique & SoC \newline Integrator     & Foundry & Test Facility & End User \\ \midrule
    Watermarking~\cite{kahng1998watermarking}                                               & \OK & \OK & \OK & \OK \\
    Camouflaging~\cite{syphermedia, JV_2013_CCS_camo, camoStdCell}                       & \OK & \OK & \OK & \NO \\
    Split manufacturing~\cite{jarvis2007split, ImesonSEC13}                                 & \OK & \NO & \OK & \OK \\
    Metering (passive)~\cite{hardwaremeteringusenix, koushanfar2012provably}                & \OK & \OK & \NO & \NO \\
    Logic locking~\cite{roy_epic:_2008,fll,sensitization,sarlock,antisat,sfllhd,sfll-rem} & \NO & \NO & \NO & \NO \\ \bottomrule
    \end{tabular}%
    }
    \aftertable
    \end{table}
    
\begin{figure*}[tb]
    \centering
    \includegraphics[width=\textwidth]{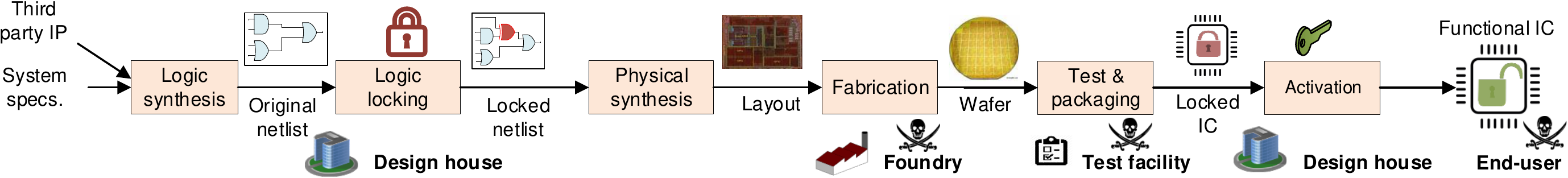}
    \caption{The IC design flow: this involves a geographically distributed supply chain with numerous untrusted parties.}
    \label{fig:design-flow}
    \afterfig
    \end{figure*}

    The IC supply chain is geographically distributed and involves potentially untrusted parties that expose IP to piracy threats~\cite{rostami_primer_2014,IEEE_proceedings_Counterfeits}. 
    Potential adversaries in the design flow include System-on-Chip (SoC) integrators, foundries, test facilities, and end-users, as shown in~\autoref{fig:design-flow}. 
    Throughout the design flow, there are opportunities for overbuilding, reverse-engineering, and hardware Trojan insertion~\cite{xiao_hardware_2016}. 
    Researchers have proposed various digital circuit "Design-for-Trust" techniques, including watermarking~\cite{kahng1998watermarking}, camouflaging~\cite{syphermedia, JV_2013_CCS_camo, camoStdCell}, split manufacturing~\cite{jarvis2007split,ImesonSEC13}, passive metering~\cite{hardwaremeteringusenix,koushanfar2012provably}, and logic locking~\cite{roy_epic:_2008, RLL, sfll-rem, chakraborty2009harpoon, ttlock, rezaei-cyclic}, as shown in~\autoref{tab:intro-table}.
    
\begin{figure}[b!]
    \centering
    \includegraphics[width=0.8\columnwidth]{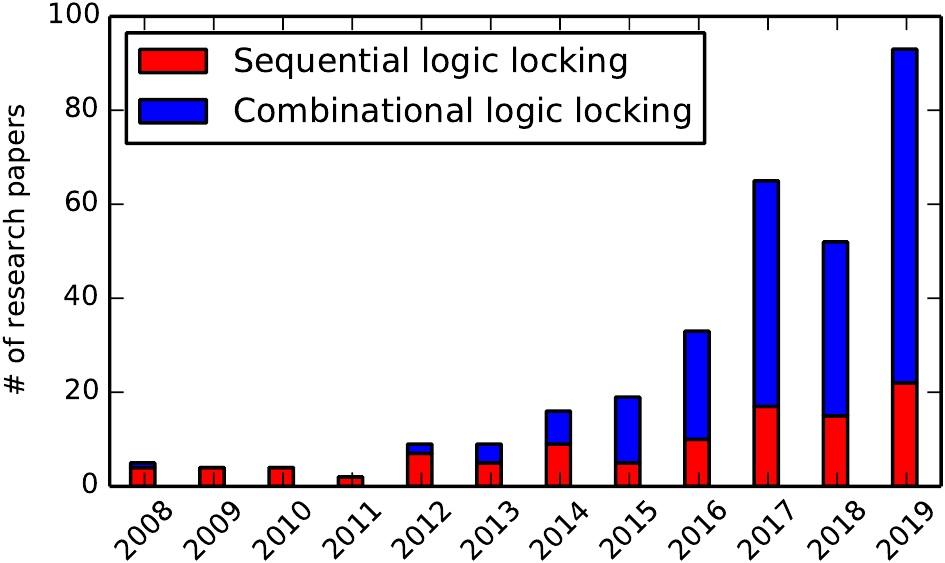}
    \caption{Publications in logic locking over the years.}
    \label{fig:publication-trends}
    \afterfig
    \end{figure}

    We focus in this work on \textit{logic locking} as a promising and versatile approach for protecting IP against various threats in the supply chain; it has gained traction in the wider hardware design academic community (reflected in the trajectory of publications in~\autoref{fig:publication-trends}). 
    Logic locking is an active, decade-old research area~\cite{Yasin2020}, with numerous proposals (e.g.,~\cite{roy_epic:_2008, RLL, sfll-rem, chakraborty2009harpoon, ttlock, rezaei-cyclic}) and subsequent attacks (e.g.,~\cite{sat-attack,appsat_host_2017,sensitization,bitflip,shen2017double,cycsat,redundancy_attack}). 
    Given logic locking's status as a frontier of hardware security, this paper contributes: (i)~a critique on logic locking evaluation in the literature, (ii)~community-driven benchmarking to provide a "snapshot" of the current field, and (iii)~broader lessons for hardware security benchmarking. This work systematizes an important area of IC supply chain security, and as other hardware security techniques approach maturity, these experiences offer useful insights for future benchmarking efforts. 
    
\subsection{What is Logic Locking?}

    Broadly, IP usurpers want to take a design and use it illegitimately. 
    As illustrated in~\autoref{fig:design-flow}, digital designs undergo a series of processes. 
    Within a \textit{design house}, designers integrate various IPs to satisfy a system specification; this design undergoes logic synthesis to produce a netlist. 
    The netlist is transformed through physical synthesis to produce a layout that can be fabricated by a \textit{foundry}, producing the IC. 
    After manufacture, the ICs are sent to a \textit{test facility} for testing and packaging, before being returned to the design house. Logic locking protects a design by inserting additional logic into it; the added logic "locks" the design such that a locked circuit, without the correct key, behaves erroneously. 
    To protect combinational circuits, logic locking transforms the output into a function of the primary and the key inputs. To protect sequential circuits, logic locking inserts states and state transitions, expanding the reachable state space to include non-functional states.  The netlist is locked in the \textit{design house} before being sent to the other parties in the design flow; the locked IC is \textit{activated} by the \textit{design house} before distribution to \textit{end-users}. 
    Logic locking aims to mitigate several threats: 
    
\begin{itemize}[leftmargin=*,noitemsep]
\item     An untrusted foundry or an end-user may \textbf{reverse engineer and pirate IP} by extracting a netlist from the physical IC or layout. Without the key, an attacker cannot produce a design that is functionally equivalent to the original.
\item  A malicious foundry may \textbf{overbuild} surplus ICs and sell them in grey markets at lower prices. However, without the secret key, the foundry cannot unlock the extra ICs. 
\item \textbf{Hardware Trojans} are malicious alterations to a design. A logic locked IC can prevent insertion of Trojans. The key gates alter the transition probabilities of the signals in a circuit in a manner unknown to the attacker, making it harder for them to identify stealthy locations for Trojans.
\item Logic locking can hamper {\bf counterfeiting} and cloning, as they rely on reverse engineering~\cite{IEEE_proceedings_Counterfeits}.  
\end{itemize}

\subsection{What is the current landscape in logic locking research?}
While logic locking is touted by academic proponents as a promising defense against IP piracy, reverse-engineering, and hardware Trojan insertion~\cite{dupuis_novel_2014}, the community at-large continues to contribute numerous back-and-forth exchanges, extolling the latest triumphs against attacks or proclaiming the "end of logic locking"~\cite{engels_end_2019, rahman_key_2019}. 
Throughout the recent, formative years of logic locking research, various independent research groups have investigated the effectiveness of attacks and defenses alike. 
This raises a natural question: \textit{What is the current landscape in logic locking research}? 
To answer this question, we systematize the literature and identify several shortcomings in how logic locking techniques have been evaluated.
As we will explain in~\autoref{sec:preliminaries}, these include: 
\begin{itemize}[leftmargin=*,noitemsep]
    \item Reliance on "best-effort" re-implementation of techniques
    \item Inconsistent locking parameters for comparisons
    \item Variations in comparison criteria (e.g., allowed attack time)
    \item Inconsistent benchmark selection
    \item Limited benchmark complexity 
    \item Disagreement over "in-scope" security guarantees
\end{itemize}
Motivated to overcome these deficiencies, we sought inspiration from cybersecurity contests across software and hardware domains~\cite{nist_csrc_2017, ctf-list, fasano_rode0day_2019, cyberchallenge, csaw-esc, dessouky_hardfails:_2019, clavier_practical_2014} as a means to synthesize community insights into hardware security.  We highlight examples of such contests in~\autoref{tab:other-contests}, where the exercises brought together the community to explore new approaches, benchmark techniques, and systematize knowledge. Besides synthesizing insights from the literature, we embarked on the first community-wide benchmarking of logic locking. This took the form of a red-team/blue-team "competition" with a common framework for evaluating the locking techniques and the efficacy of attacks on logic locking. By bringing together leading research groups with interested experts from government agencies and the private sector, this effort aimed to build a capability towards something akin to a NIST-style contest (e.g., post-quantum cryptography process~\cite{nist_csrc_2017}).

\begin{table*}[t]
\centering
\caption{Selected Community Competitions in Cybersecurity-related domains\label{tab:other-contests}. For "Open Source": \Circle~indicates that the contest involved/expected attacks and defenses to be open sourced, \CIRCLE~indicates that there was no expectation for either attack or defense to be open, \RIGHTcircle~indicates that the attack artifacts were open-source but not the defense}
\resizebox{\textwidth}{!}{%
\begin{tabular}{@{}L{2cm}L{1.2cm}C{1cm}C{1.2cm}L{2.6cm}L{1.5cm}L{9cm}@{}}
\toprule
Competition (First year) & Organizer & Open Source & Domain & Aims & Participants & Contributions \\ \midrule
Crypto standards (e.g., AES, PQC) (1997)  & NIST & \Circle & Crypto & Standardize crypto algorithms & $\geq$ 300 & Participants submit candidates for evaluation. Community contributes to minimum requirements, submission requirements, evaluation~\cite{nist_csrc_2017} using open comment threads and workshops \\  \cmidrule(r){1-7}
CTF (1996) & Academia+ Govt.& \CIRCLE & Software &  Vulnerability discovery and exploitation & $\geq$ 10,000 annually & Contestants identify and exploit system vulnerabilities. These contests are offered as professional development opportunities, community-building exercises, and product security evaluation~\cite{ctf-list}\\ \cmidrule(r){1-7} 
Rode0day (2018) & Academia+ Govt. &\RIGHTcircle & Software &  Automated Binary Bug Detection & $\sim$ 28 teams & Monthly "buggy binary" released. Participants aim to discover vulnerabilities. Bug injection is automated using an open-source tool~\cite{fasano_rode0day_2019}\\ \cmidrule(r){1-7}
Cyber Grand Challenge (2014) & DARPA & \Circle & Software &  Autonomous software vulnerability detection and patching & $\sim$ 200 &  Participants develop and test autonomous systems to identify vulnerabilities and develop patches for unknown software~\cite{cyberchallenge}. The organizers prepared the challenge sets with vulnerabilities for the competitors' automated tools to analyze and repair. Tools generated proof of vulnerabilities and deployed them against competing tools. \\ \cmidrule(r){1-7}
CSAW-ESC (2008) & Academia  & \Circle & Hardware &  Hardware security & $\geq$ 300& Contest spanning numerous domains, including hardware Trojan design, detection~\cite{csaw-esc}, yielding the  hardware Trojan corpus~\cite{trusthub}      \\ \cmidrule(r){1-7}
Hack@DAC (2017) & Academia+ Industry  & \RIGHTcircle & Hardware & Hardware Bug Detection (CTF-style) & $\geq$ 150& Participants identify bugs in an SoC. They score points by submitting bug reports identifying the root cause, a test for confirming the issue, severity assessment, and exploit code. They may submit tools for automatic bug detection. Offers insights into the challenges for manual/automated bug detection in hardware flows~\cite{dessouky_hardfails:_2019}\\ \cmidrule(r){1-7}
DPA Contest (2008) & Academia  & \Circle & Hardware/ Crypto &  Side-channel resistance & $\geq$ 30 & Compare side-channel techniques (attacks, acquisition techniques, counter-measures). Participants develop attacks and defenses, organizers prepare benchmarks and draw insights into sample acquisition challenges~\cite{clavier_practical_2014}\\ \hline \hline
This work (2019) & {Academia} & \CIRCLE & {Hardware}  & {Logic Locking Evaluation} & {$\sim$ 40} & {Evaluate state-of-the-art combinational and sequential logic locking against numerous attacks to draw insights from participants' experiences}\\ \bottomrule
\end{tabular}%
}
\aftertable
\end{table*}

\textbf{Contributions.} %
From the combined efforts of the community, this study offers new insights for evaluating hardware security techniques. The contributions include:
\begin{itemize}[leftmargin=*,noitemsep]
    \item A critical analysis of the evaluation of attacks and defenses in the literature. 
    \item Community-guided insights into threat models and attacker capabilities addressed by logic locking.
    \item Results from red-team/blue-team benchmarking. For the first time, red teams could launch attacks on a common set of benchmarks, allowing a fairer comparison of their efficacy and a measure of the resilience of the blue teams' defenses. 
    \item Insights and reflection on the overall benchmarking process, including open questions and opportunities for investigation.
\end{itemize}

\subsection{Paper Organization} %
In~\autoref{sec:preliminaries}, we outline the broad aims of logic locking and present our critical analysis of the logic locking literature. 
In~\autoref{sec:invasive}, we address recently proposed \textit{invasive} attacks on logic locking. 
Following this, we discuss our benchmarking effort, discussing the threat model and assumptions on attacker capabilities in~\autoref{sec:threatmodel}. 
In~\autoref{sec:defenses}, we provide brief technical background on the combinational and sequential locking techniques and then present the findings from the benchmarking effort in~\autoref{sec:csaw}, where we outline the various attack approaches and the achieved attack results. 
In~\autoref{sec:discuss}, we reflect on our efforts, examining lessons learned. 
Finally, we present our perspectives on the future of logic locking and its evaluation, concluding in~\autoref{sec:future}. 

\section{Critical Analysis of the Literature}
\label{sec:preliminaries}

\begin{table*}[tb]
\centering
\caption{Non-exhaustive overview of defense/attack self-evaluation: Circuits Used, Techniques (Re-)Implemented, and Evaluation Design}
\label{tab:evaluation-approaches}
\resizebox{\textwidth}{!}{%
\renewcommand{\arraystretch}{0.1}
\begin{tabular}[t]{@{}L{0.1cm}L{0.3cm}L{2cm}L{3.6cm}L{3cm}L{10cm}@{}}
\toprule
& Year & Technique     & Evaluation Circuits \newline (number in chosen subset) & (Re-)implemented Locking/ Attacks  & Notes on Evaluation Design (e.g., key sizes)  \\ \midrule
& 2008 & RLL~\cite{RLL}  & ISCAS85 (2)& -- / -- & Informal/Qualitative security analysis focusing on the key distribution aspects \\ \cmidrule(r){2-6}
\multirow{8}{*}{\rotatebox{90}{Combinational Locking Techniques}} & 2012 & FLL~\cite{fll} & ISCAS85 (10)& -- / -- & Informal/Qualitative security analysis \\ \cmidrule(r){2-6}
& 2012 & SLL~\cite{sensitization}            & ISCAS85 (9)& RLL & Brute-force analysis, key gate-type analysis, \# of test pattern analysis, informal analysis \\ \cmidrule(r){2-6}
& 2016 & SARLock~\cite{sarlock} & ISCAS85 (4), OpenSPARC (7)   & RLL, SLL / SAT & Start of the "post-SAT" techniques, applies SAT from~\cite{sat-attack}, \textbf{10--64} key-bits \\ \cmidrule(r){2-6}
& 2016 & Anti-SAT~\cite{antisat} & ISCAS85 (3), MCNC (3) & SLL / SAT & SAT from~\cite{sat-attack} with 10 hr timeout, \textbf{43--364} key bits, SLL based on \textbf{5\%} area overhead \\ \cmidrule(r){2-6}
& 2017 & TTLock~\cite{ttlock}        & ISCAS89 (5)                  & -- / SAT, Sensitization & \textbf{16--18} key bits, used SAT~\cite{sat-attack}, 48 hr timeout \\ \cmidrule(r){2-6}
& 2017 & Cyclic~\cite{cyclic}        & ISCAS85 (9), MCNC (7) & RLL (for area comparison only) / -- & \textbf{72} key-bits to compare overhead, no empirical attack evaluation \\ \cmidrule(r){2-6}
& 2017 & SFLL-HD/ SFLL-flex~\cite{sfllhd}       & ISCAS89 (3), ITC99 (7), ARM Cortex-M0 & FLL / SAT, Anti-SAT, 2-DIP, Sensitization & 48 hour timeout for SAT, \textbf{11--14} key-bits for initial evaluation, \textbf{128} key-bits of SFLL and \textbf{128} key-bits of FLL for case study on ARM Cortex-M0 \\ \cmidrule(r){2-6}
& 2018 & SFLL-fault~\cite{sfll_fault}   & ITC99 (6) & -- / SAT & \textbf{128} key-bits, 48 hour timeout. Claims the same as SFLL-flex \\ \cmidrule(r){2-6}
& 2019 & SFLL-rem~\cite{sfll-rem} & ITC99 (6), DARPA CEP SoC, ARM Cortex-M0 & -- / SAT, FALL & \textbf{80} key-bits, evaluated using open-source implementations from~\cite{sat-attack} and~\cite{fall}, \textbf{524} key-bits for SoC case study  \\ \midrule \midrule
& 2012 & Sensitization~\cite{sensitization} & ISCAS85 (10) & RLL, FLL / -- & Brute-force analysis, key gate-type analysis, \# of test pattern analysis \\ \cmidrule(r){2-6}
\multirow{8}{*}{\rotatebox{90}{Attacks on Combinational Locking}} & 2015 & SAT~\cite{sat-attack} & ISCAS85 (11), MCNC (12) & RLL, FLL / Sensitization & Released a SAT-based attack tool and tool for applying RLL, FLL, and SLL, set 10 hour timeout. Locking based on \textbf{10--50\%} overhead. \\ \cmidrule(r){2-6}
& 2017 & SPS~\cite{sps} & ISCAS85 (8) and OpenSPARC (7) & FLL, Anti-SAT / SAT & Compares SAT attack to SPS, \textbf{64 \& 256} Anti-SAT key-bits + FLL based on \textbf{5\%} area overhead \\ \cmidrule(r){2-6}
& 2017 & AGR~\cite{agr} & ISCAS85 (10), ISCAS99 (2) OpenSPARC (7), MCNC (3)  & FLL, Anti-SAT, SARLock / SPS, AppSAT & \textbf{512} Anti-SAT key-bits + FLL based on \textbf{5\%} overhead \\ \cmidrule(r){2-6}
& 2017 & Bypass~\cite{bypass} & ISCAS85 (6), MCNC (1), EPFL (1) & RLL, SARLock, Anti-SAT & Informal comparison of attack with SAT and Removal on different techniques, mentions SLL but implements RLL for SARLock+SLL \\ \cmidrule(r){2-6}
& 2017 & AppSAT~\cite{appsat_host_2017} & ISCAS85 (20), MCNC (6) & Anti-SAT, RLL  / -- & \textbf{16--30} Anti-SAT key-bits + \textbf{30--80} RLL key-bits \\ \cmidrule(r){2-6}
& 2017 & 2-DIP~\cite{shen2017double} & MCNC (11) & SARLock, ~\cite{dupuis_novel_2014} / SAT & \textbf{8--256} SARLock key-bits with \textbf{27--1682}~\cite{dupuis_novel_2014} key-bits, representing \textbf{5--25\%} area overhead\\ \cmidrule(r){2-6}
& 2017 & CycSAT~\cite{cycsat} & ISCAS85 (9), MCNC (12) & \cite{dupuis_novel_2014} using~\cite{sat-attack}, Cyclic / SAT & \textbf{71--90} key-bits, uses~\cite{dupuis_novel_2014} based on 10\% area overhead, \textbf{53--837} key-bits total; illustrates shortcoming of~\cite{cyclic,sat-attack} as a result of specific SAT-attack implementation\\ \cmidrule(r){2-6}
& 2019 & Redundancy~\cite{redundancy_attack} & ISCAS85 (10) & RLL, Sensitization / -- & Oracle-less individual and pair-wise analysis based on \textbf{5\%} area overhead \\ \cmidrule(r){2-6}
& 2019 & Unit Function~\cite{zhang2019tga} & ISCAS85 (4), ITC99 (6) & RLL / -- & Oracle-less self-referencing search of equivalent unit functions for \textbf{128} key bits \\ \cmidrule(r){2-6}
& 2019 & FALL~\cite{fall} & ISCAS85 (9), MCNC (11) & TTLock, SFLL-HD / -- & SFLL-HD$^{0}$, SFLL-HD$^{h}$ with \textbf{10--128} key-bits, 1000 sec. timeout\\ \bottomrule
\end{tabular}%
}
\aftertable
\end{table*}

\subsection{Background: Motivations for Logic Locking}
Motivation for logic locking was first presented in~\cite{RLL,roy_epic:_2008}, where the primary issue is that of integrated circuit (IC) or intellectual property (IP) piracy. In this scenario, a malicious party steals designs so that they can produce or distribute them for their own benefit. As such, the main objective of~\cite{roy_epic:_2008} is to discourage "unauthorized excess production and stolen masks" by making it difficult to reverse-engineer or modify a functional IC. 
This motivation has since inspired the research community, giving rise to threat models  for both combinational and sequential circuit locking research.  The intended users of logic locking include semiconductor IP designers and system integrators who want to protect high-value designs from theft and manipulation by untrusted foundries and from the threat of reverse-engineering. 

\subsection{Logic Locking Evaluation in the Literature}
\autoref{tab:evaluation-approaches} presents an overview of work in combinational locking, highlighting techniques and settings used by authors in evaluating their locking techniques or attack approaches. We invite interested readers to examine~\cite{Yasin2020} or~\cite{chakraborty_keynote_2019} for a thorough survey of the history of logic locking and related fields. 
While we do not present sequential approaches here, readers can observe similar ad hoc and disparate evaluations in the literature. 

In the initial proposals for logic locking, starting with Random Logic Locking (RLL), as part of EPIC~\cite{RLL}, techniques are evaluated on generally informal, qualitative criteria, based on novel aspects, such as number of test patterns for sensitizing primary outputs to key inputs~\cite{sensitization}. 
The formulation of a boolean Satisfiability (SAT)-based attack~\cite{sat-attack} provided a watershed moment---not only did they present an attack that was able to "defeat" existing logic locking techniques, they also provided a freely usable attack/defense implementation. 

Spurred by the SAT-based attack, new SAT- resistant techniques have proliferated, each applied to varying subsets of ISCAS '85 circuits~\cite{hansen_unveiling_1999} and circuits sourced from other benchmarks (such as MCNC~\cite{mcnc}, OpenSPARC~\cite{opensparc}, and ITC '99~\cite{basto_first_2000}). Different works chose different subsets (often with overlaps), but with different re-implementations of prior techniques, with different settings. 
For example, the choice of "how much locking" when preparing benchmarks for evaluating techniques varied across works; in some papers, e.g.,~\cite{antisat,agr,sps} choose a 5\% area overhead for guiding locking evaluation---others, such as~\cite{sfllhd}, simply picked a key size (such as 128 bits). 
Following the dissemination of~\cite{sat-attack}, the number of key bits that were successfully retrieved (in a given time) became frequent evaluation criteria for defenses, resulting in various time limits for running attacks, ranging from seconds (set in~\cite{fall}) to days (set in~\cite{sfll_fault}).
From examining the variability in evaluation design, several questions arise about the present state of logic locking attack/defense evaluation:
    \begin{itemize}[leftmargin=*,noitemsep]
        \item What is the right key size?
        \item Are oft-used benchmarks (like ISCAS) representative of "typical" IPs that logic locking might protect?
        \item What is a reasonable attack timeout?
        \item What are "realistic" capabilities of attackers? (we examine this more specifically in~\autoref{sec:invasive}) 
    \end{itemize}
The answers to these questions differ across studies, which brings to question the generalizability of claims about whether locking is "provably secure"~\cite{sfllhd}, will "most likely never succeed"~\cite{engels_end_2019}, or somewhere in-between. We take this uncertainty as an opportunity for a coordinated evaluation. 
    
\textbf{What is the right key size?} Various papers set the "amount" of locking using some holistic measure---usually as part of some security/power/performance/area trade-off (e.g., locking based on 5\% overhead as in~\cite{agr}). 
In other works, key size is often set independently of the target circuit characteristics, (e.g., in~\cite{fall}, where key size is set to a maximum 128 bits). 
When key size is set by researchers for their attack evaluations, it is hard to know if the re-implementation settings are aligned with defense designers' intentions. 
Furthermore, while there may be some overlap between the circuits used for evaluation, there is not always a 1:1 correspondence, which makes comparisons somewhat inconclusive. 

\textbf{Are ISCAS '85 benchmarks enough?} The most-used benchmark circuits come from the ISCAS '85 collection~\cite{hansen_unveiling_1999}, featuring circuits from about 6 to 3000 gates with dozens to hundreds of inputs and outputs. 
Even where ISCAS '85 circuits are used, the selection of circuits within the benchmark set varies from work to work. 
In addition, these designs are small compared to "realistic" designs, such as the IPs used in SoCs (e.g., in~\cite{sfll-rem}, the IPs are around 16$\sim$156K gates, orders of magnitude larger than the ISCAS '85 circuits).
As the ISCAS '85 circuits are not necessarily representative of the scale of complex modern IPs that will benefit from logic locking techniques, whether the claims about the scalability of locking (and corresponding attacks) hold for "realistic" scenarios requires further examination. 
While not yet widespread, recent works~\cite{sfll-rem, menon_system-level_2019} have started using realistic SoCs, such as the Common Evaluation Platform~\cite{cep} which has a UCB Rocket-Chip and integrated  DES3, GPS, MD5, RSA, SHA256, DFT, IDFT, and IIR cores connected by a AXI4-Lite bus.

\textbf{What is a reasonable attack time out?} 
Similar to the variability in key size and circuit selection, quantifying attack resilience/efficacy also varies from study to study. 
The community is evaluating security by determining what is "practical", at least empirically.
In some locking technique studies, authors aim for resilience to the SAT-based attack~\cite{sat-attack} (e.g.,~\cite{antisat,ttlock}) as measured by whether the SAT-based attack times out by taking longer than a pre-specified bound. As attack time is influenced by many factors such as the circuit complexity and key sizes and not determined solely by the attack algorithms, another criterion may be complexity in terms of the number of iterations needed to recover a key. 
There are no guarantees that the underlying execution platform resources are consistent across evaluations, and no contextualization/justification as to what constitutes a "reasonable" amount of time that a dedicated attacker would be willing to devote before an attack becomes uneconomical. 
Another missing piece of the puzzle is whether "attack time" should consider the wider effort required of an attacker; for example, should the time taken to setup tools, understand the benchmark formats, and so forth, be a factor in a holistic evaluation? 

\subsection{Shortcomings in Evaluations: A Critical View }
As evident in recent work, there is a lack of agreement with respect to metrics for the trade-off between security and overhead~\cite{hu_security-driven_2019,
menon_system-level_2019}. 
As such, there are shortcomings in the framing, analysis, and evaluation of attacks and defenses, including:

\textbf{Relying on "best-effort" re-implementation of techniques.} Evaluation of attack and defense techniques are somewhat ad hoc in the literature; independent groups typically select what they find to be the state-of-the-art attack/defense techniques, applying them on relatively small scale designs (either bespoke, or drawn from benchmarks designed for other purposes). Due to a lack of coordination or implementation availability, the techniques are often re-implemented from (often incomplete) descriptions found in literature. 

This reduces transparency of the evaluation process. While techniques are somewhat scrutinized (and there is a case to be made on the merits of replicability), there are potential biases at play (for example, choice of circuits and key sizes) and limitations on the quality of the (re-)implementations used to compare techniques. Furthermore, only some logic locking attack and defense implementations are open-sourced.  To improve attacks and defenses, attackers need feedback on how well their techniques break locking, and defenders need feedback on how well their techniques perform under scrutiny.

\textbf{Inconsistency in setting parameters for comparisons.} In addition to the re-implementation of techniques, attack and defense techniques feature several parameters. 
These include which combination of techniques to apply, the number of key bits, which attack techniques to compare with, and attack time-out limits. 
Given the wide variety of settings, direct comparisons between different techniques are difficult to make, so conclusions related to relative performance/overhead impacts vs. attack success, etc. do not necessarily provide clear guidance to practitioners as to appropriate key sizes or security/overhead trade-offs. 
Moreover, different attacks are performed with varying computational resources, from student desktop-class machines through to university high-performance clusters, further complicating direct comparisons.

\textbf{Variations in comparison criteria.} When comparing locking and attacking techniques, researchers use  numerous criteria to compare their works to prior art. These include area overhead, performance impact, and resilience against a chosen attack (e.g., measured as time taken to perform a SAT-based attack~\cite{sat-attack}, or a specific \textit{implementation} of a SAT-based attack).  On this front, individual contributions to the literature are somewhat scattershot: some works might trade-off area overhead against number of SAT-based iterations, whereas others might use analytic measures (e.g., theoretical corruptibility). Different groups choose different evaluation criteria, while making broader claims about scalability, attack resiliency, or attack effectiveness against types of defenses.  As observed in~\cite{menon_system-level_2019}, there are no unified metrics in terms of circuit size, SAT-attack execution time, or technology libraries.

\textbf{Limited benchmarks.} As mentioned earlier, numerous works use a limited set of evaluation circuits, typically drawn from the ISCAS '85 benchmarks~\cite{hansen_unveiling_1999}, 
ITC '99 set~\cite{basto_first_2000} and the MCNC (ACM/SIGDA) benchmark suite~\cite{mcnc}.
While they may not match the complexities of IPs in modern systems-on-chip (SoCs), they provide valuable insight into how locking techniques change logical structure and affect overhead. Nevertheless, they may miss important issues/flaws that can arise when applying logic locking to large modern designs. While these oft-used benchmarks provide some commonality across works in the literature, they provide limited range in terms of reflecting different applications and use cases, especially as the community moves forward towards developing more useful and robust techniques. 

Furthermore, with respect to emulating attack scenarios, the use of known benchmarks provides advantages to an attacker (beyond the scope of the threat model). Additionally, ISCAS and ITC benchmarks were developed by the IC test community for evaluating automatic test pattern generation (ATPG) coverage.  Other research communities, such as those working on hardware security, have adopted these benchmarks. Crucially, for logic locking, these benchmarks do not have verification testbenches to prove that locked circuits behave identically to original circuits or that a key has correctly been applied. Another practical issue is that these benchmarks are often in \texttt{BENCH} format (from the ISCAS benchmarks~\cite{hansen_unveiling_1999}), which is not easily compatible with several EDA tools. 

\textbf{Limited resources for coordinating.} Aside from the effort we present in this paper, there have not been many opportunities for a red team/blue team style evaluation of attacks and defenses, where an impartial party facilitates the interaction between participants and allows organized sharing of resources (e.g., benchmark circuits and external judging).
Without this coordination, the barrier of entry into the logic locking attack/defense domain can be prohibitively high, as teams need to prepare both attack techniques and evaluation artifacts. 
Running an exercise that frees teams from the need to re-implement allows attackers and defenders to focus on their techniques; this also enables a controlled and fair comparison across techniques (and their implementations). 

\textbf{Disagreement over "in-scope" security guarantees.} There are myriad threat models that aim to be addressed by myriad defense techniques.
In the existing literature, defender goals (when proposing a locking mechanism) and attacker points-of-view (in terms of what constitutes the "breaking" of a technique) do not always align. 
For example, the work in~\cite{sfll-rem} assumes an attacker that can reverse-engineer a netlist from a physical layout (e.g., GDSII files) and also acquire a working chip (e.g., from the "open-market"), setting aside invasive attacks as orthogonal (and addressed by advances in physical, package-level countermeasures). 
In contrast, recent critiques of logic locking~\cite{engels_end_2019,rahman_key_2019} propose invasive attacks, such as optical probing. 
Such attacks do not identify weaknesses in the fundamental strength of the proposed locking techniques, but instead target the underlying implementation technologies---these vulnerabilities can be addressed with suitable technology-based solutions that build on top of logic locking in a defense-in-depth approach~\cite{rahman_defense--depth_2020}. 
We discuss more challenges to invasive attacks in~\autoref{sec:invasive}.

\subsection{Informing our Benchmarking Exercise}
Given the plethora of evaluation settings, we were motivated to bring the community for a coordinated benchmarking effort.
In other words, the aim of this benchmarking exercise was to provide a level playing field for comparing different attacks against locking techniques, allowing teams to field their "best efforts" by studying a common set of circuits within a common time-frame. Our benchmarking exercise was thus designed to address these shortcomings in the current landscape by:

\begin{itemize}[leftmargin=*,noitemsep]
\item agreeing on a threat model for the benchmarking exercise (see \autoref{sec:threatmodel});
\item setting common goals to compare attack techniques (i.e., key recovery, see \autoref{sec:threatmodel});
\item preparing a shared benchmark suite with clearly defined locking parameters (see \autoref{sec:defenses}); 
\item focusing on larger, realistic benchmarks  (see \autoref{sec:defenses});
\item empowering the teams to focus on their attacks/defenses, without fretting about re-implementing others' works (see \autoref{sec:csaw}); and 
\item coordinating the community's effort (see \autoref{sec:csaw}).
\end{itemize}

\section{On Invasive Attacks and Assumptions\label{sec:invasive}}
In~\cite{engels_end_2019} and~\cite{rahman_key_2019}, authors proposed new threat models based on the availability of advanced probing techniques such as e-beam~\cite{schlangen_backside_2007} and laser probing~\cite{boit_ultra_2013,kindereit_fundamentals_2014} that were developed for IC failure localization analysis. 
Although spontaneous light emission techniques~\cite{song_marvel_2011,song_advanced_2005} were not mentioned in~\cite{engels_end_2019,rahman_key_2019}, they have also successfully been used to probe internal IC nodes. 
These probing tools were originally invented and developed for IC failure analysis and chip characterization~\cite{boit_ultra_2013,song_advanced_2005}. 
Typically, they are used for defect localization through global inspection (using light emission maps and electro-optical frequency mapping) and then local probing of the specific nodes. 
Usually these techniques are time-consuming and require preliminary electrical test and diagnostics aimed to narrow down the probing areas. 

These tools have been expanded and used for hardware security by developing automated data collection, image processing, and information extraction~\cite{song_marvel_2011,stellari_revealing_2016}. 
Some applications include identifying functional blocks and detecting Trojans~\cite{song_marvel_2011}, and reading the contents of 14~nm SRAM~\cite{stellari_revealing_2016}. 

Fundamentally, there are two issues when attackers want to use optical probing techniques for observing the key gate signals or key latches/flip-flops. 
One is the limitation of spatial resolution, and the other is identifying the probing location. 
As the technology evolves to 7~nm and below, it becomes very difficult for any non-invasive optical probing tool to be able to see individual gates, even using Solid Immersion Lenses (SILs)~\cite{mansfield_solid_1990}. 
The optical spatial resolution of a state-of-the-art probing tool equipped with SIL is about 200~nm which can be calculated using the Rayleigh criterion R=0.61$\lambda$/NA, where $\lambda$ is the light wavelength (e.g., 1064~nm or 1340~nm for commonly available probing lasers), and NA is the SIL numerical aperture (e.g., $\sim$3). 
The same principle can be applied to the resolution of transistors that emit light emissions when they are powered up or switching. 
These emissions are broadband and reach a peak around 1.8 µm for new technology nodes~\cite{kindereit_near_2012}. 
Therefore, using light emission for probing also suffers of resolution issues.
However, even with these limitations and challenges, light emission can be used to detect some types of Trojans~\cite{song_marvel_2011} due to advances in image processing and analysis techniques. This is because the physical size of Trojans is large enough so that optical probing can identify additional circuitry compared to the original design. 

For extracting the logic locking key, however, one needs to identify the gates used in locking---this is much smaller compared to hardware Trojans. 
In~\cite{rahman_key_2019}, a 28~nm FPGA was used to demonstrate that the key values can be extracted using laser probing. 
Although this proof-of-concept example shows the success of laser probing, it will suffer when measuring random logics manufactured in the latest technology node. 
The diffraction limitation fundamentally defines the resolution of optical probing, as mentioned earlier.  For scaled nodes, the electro-optical spots are often a cluster of many gates and it is very hard or impossible to interpret them. 
It is even harder to derive the logic key values from these spots.  For random logic, designers can intentionally hide the key gates using a more compact layout. From the semiconductor process side, designers can choose heavily doped silicon wafer so that the near infrared light is absorbed by the silicon. This makes it hard to see through the backside. Therefore, key bits can be hidden without using design tricks as an anti-tamper feature. 

Localizing the region of the chip containing the locking logic is another challenge. 
In contrast to regular chip diagnostics where test engineers have some targeted test patterns and responses to guide where to probe, designers of locked circuits could make a design where keys are hard to find by probing. Although recent works~\cite{engels_end_2019,rahman_key_2019} assume that a malicious foundry can reverse engineer the IP core from the GDSII file and identify the key gates and registers, it is in fact a challenge to localize them, in particular for random logic. Together with the limits of spatial resolution at the advanced nodes, it is challenging to reveal the key bits using probing.

{\it Fundamentally, the goal of logic locking is to protect IP during manufacture.  Protecting IP after manufacture and during use is the province of anti-tamper technologies.
Both technologies must be used in a coordinated fashion to protect a chip throughout its life cycle.  
Ideally they would be complementary approaches that make both the locking and the anti-tamper stronger than they would be in isolation. 
This is a promising area for further study.}

\section{Benchmarking: Threat Model and Goals}
\label{sec:threatmodel}

\subsection{Wider Attacker Motivations}
    Following our critical examination of the literature, we now turn our attention to benchmarking. 
    In discussing any potential security problem, it is important to characterize potential attackers' motivations, targets, and capabilities. 
    Rajendran \textit{et al.} first described the potential threat posed by an attacker that can (1) reverse-engineer a locked netlist, and (2) access a functional, unlocked design (i.e., acquired from the open market)~\cite{sensitization}. 
    Access to a reverse-engineered netlist enables attackers to perform structural analysis of sequential~\cite{chakraborty2009harpoon} and combinational circuits~\cite{sensitization}. 
    With an unlocked netlist, attackers have an oracle, giving rise to query-based attacks, such as those that cast unlocking as a Boolean satisfiability (SAT) problem, e.g., as originally formulated by Subramanyan \textit{et al.}~\cite{sat-attack}. 
    This threat has oft-appeared in the literature, where it is posited that malicious designers, in a chip fabrication plant, are well-positioned to reverse-engineer the GDSII file of a design to obtain the locked netlist, and malicious end-users or IC testers are well-positioned to access a functional, unlocked chip~\cite{sfll-rem}. 
    Attackers that conspire and coordinate between these pre- and post-fabrication threat actors thus represent a plausible "worst-case" most-capable attack scenario that logic locking techniques seek to address. 

\subsection{Benchmarking Threat Model and Attacker Goals}    
    Using this threat of colluding malicious actors in the design flow as our starting point, red and blue teams discussed various challenge goals focused on unlocking the correct functionality of a locked design, i.e., working towards functional equivalence between an unlocked design and the corresponding attacker-recovered design. 
    Our discussions worked towards deciding what would be considered "in-scope" for our initial benchmarking effort. 
    As evident in the literature, this aim is framed as a problem of determining the secret key (or unlocking key sequence in sequential locking), as the recovery of the key entails full recovery of a locked IC's functionality. 
    
    In combinational locking, we also considered the possibility of \textit{approximate} functionality recovery---i.e., we discussed if a "partially" successful attack is meaningful, where an attacker only recovers a subset of the key or produces an approximately functional circuit. 
    A "partially unlocked" circuit would produce input/output functionality that matches some, but not all, of those from the original design. 
    The utility of a partially unlocked circuit depends on the application context: a circuit that has a low error rate, relative to the complete input/output space, could constitute a successful attack; in another case, perhaps with regards to mission critical designs, no error is tolerable at all. 
    Alternatively, it may be the case that no error in a \textit{user-defined} input/output subset is allowed in the partially unlocked circuit---whether or not the attacker has prior knowledge of the targeted design and its use depends on the broader context, such as the availability of collateral (e.g., datasheets, whitepapers, or patents).

    The benchmarking coalesced around the essential threat that logic locking thwarts, i.e., embracing the \textit{key-centric} model. The red teams set out to recover the secret key, with or without oracle access. As we re-iterate in~\autoref{sec:discuss}, partially unlocked circuits are useful in some contexts obviating recovery of the full key. 
    
\subsection{Benchmarking Scope}
    In the wider context of electronic design reverse-engineering, real-world attackers need to perform various tasks to recover knowledge about a design, depending on the design artifacts they are able to retrieve. 
    For example, attackers need to re-produce the netlist; given a netlist, an attacker might first need to perform a functional analysis to identify the components-of-interest in the design~\cite{subramanyan_reverse_2013}. 
    Given a manufactured IC, attackers first need to depackage, delayer, and carry out some form of imaging/microscopy to ascertain the circuit structure~\cite{torrance_state---art_2011}, after which the netlist can be recovered for further analysis. 
    Attackers may want to gain information from probing devices and measuring side-channels (such as power or EM emissions).  These techniques are by no means trivial, and represent a significant level of investment by them. 
    
    As logic locking provides one security modality among many possible complementary defenses, we focused the scope of our benchmarking effort to the evaluation of logic locking techniques themselves, setting aside the "upstream" challenges for recovering the netlist to begin with. 
    To that end, we assumed that a correct, locked netlist is available to the attacker, and that, where we make available an oracle, barriers to acquiring an unlocked working device are non-existent.

\section{Benchmarking: Overview and Preparation}
\label{sec:defenses}

\subsection{Participants and Timeline}

    \begin{figure}[htbp]
    \centering
    \includegraphics[width=0.9\columnwidth]{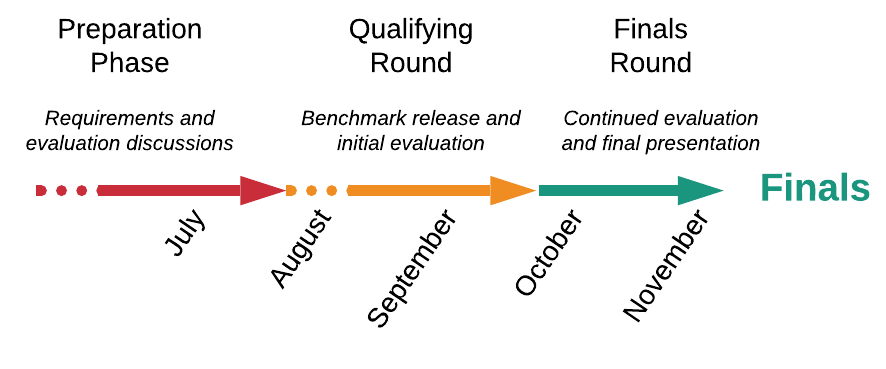}
    \afterfig
    \caption{Timeline of the Benchmarking Process}
    \label{fig:timeline}
    \afterfig
    \end{figure}

    The benchmarking involved over 40 experts registered in 18 teams from 14 affiliations in the USA and India. The teams comprised experts in VLSI test, hardware security, and academics working at the forefront of combinational and sequential logic locking research. Additionally, we invited seven external judges from industry and government agencies alongside some industry observers. They interacted with/quizzed the participants at the finals event, where the attack teams presented their approaches and outcomes. 
    
    The benchmarking effort involved two rounds, as illustrated in~\autoref{fig:timeline}. 
    The qualifying round lasted $\sim$1.5 months. The red teams attacked the locked designs created by the blue teams while maintaining a live journal on their attack methods and the tools used. At the end of this round, the red teams submitted preliminary results (i.e., key bits and/or input sequences) for feedback. Red teams with promising approaches were down-selected as finalists. The finals round lasted $\sim$1 month and involved further attack work, culminating in the delivery of a final report, poster, and in-person presentation to the judging panel. During the challenge, we invited teams to reach out to the blue teams as required (via the challenge coordinators) for clarification or technical support. There were no restrictions on the red teams with respect to compute resources or tools\footnote{a tacit expectation was that the teams did not subvert the spirit of the competition by reverse-engineering the oracle (executables).}. 

\subsection{Benchmark Preparation}
    Participants focused on combinational and sequential locking separately, with two expert blue teams preparing respective sets of locked designs. 
    Red teams had the freedom to choose which class and size of circuits to attack. 
    This section describes the locking techniques selected for evaluation and provides details on the size/complexity of the locked designs. The details are provided from the perspective of the blue teams for combinational ({\it Team X}) and sequential locking ({\it Team Y}). 

\subsubsection{Combinational Locking Overview}
\begin{figure}[tb]
\centering
\subfigure[]{\includegraphics[width=0.2\textwidth]{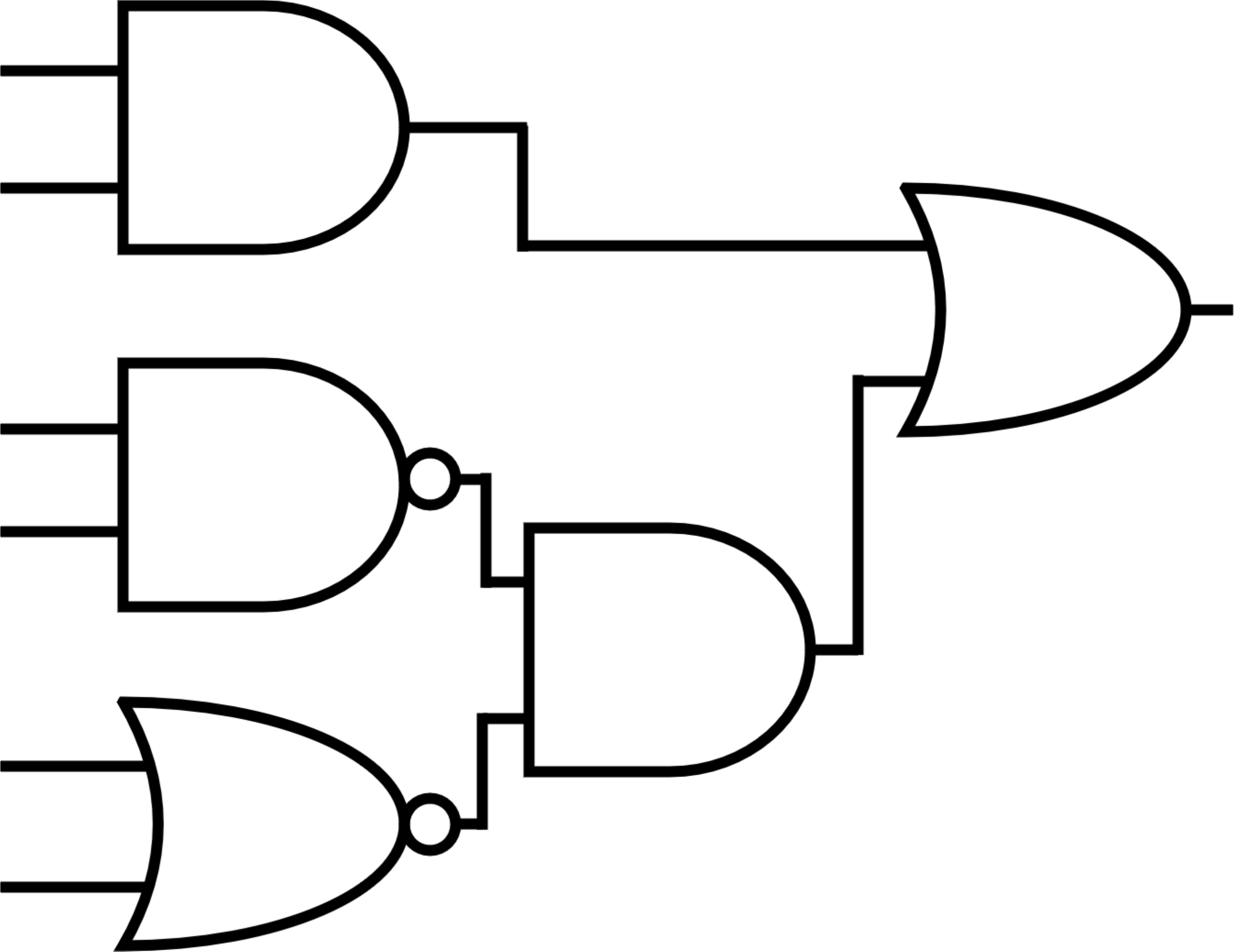}\label{fig:orig}}
\hfill
\subfigure[]{\includegraphics[width=0.27\textwidth, trim = {0cm 0.5cm 0cm 0.7cm}, clip = true]{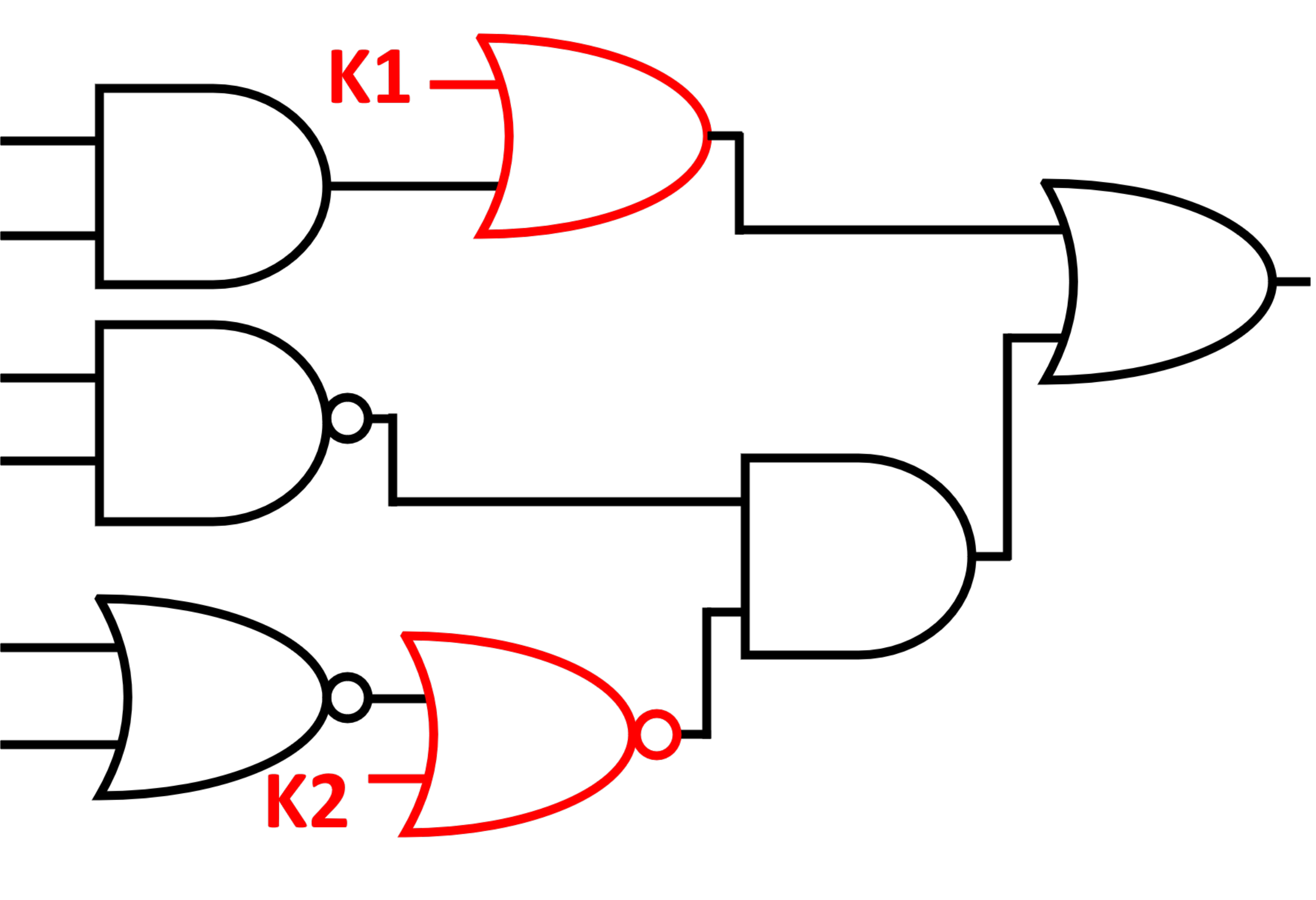}\label{fig:locked}}
\hfill
\caption{(a) Original IP (b) Locked IP with two randomly inserted key gates.}
\label{fig:locking}
\afterfig
\end{figure}

In combinational locking, the design IP is locked with key inputs by inserting some logic gates; without the correct value provided at these key inputs, the design IP will produce incorrect outputs and fail to work. The correct key value is kept secret from untrusted entities. 
\autoref{fig:locking} shows an example logic locking implementation using random logic locking (RLL)~\cite{RLL} which adds "key gates" throughout a design to corrupt the output with incorrect key inputs. In this benchmarking, {\it Team X} adopts two techniques in tandem: Stripped Functionality Logic Locking (SFLL-rem)~\cite{sfll-rem} and RLL~\cite{RLL}.
The attackers need to circumvent both layers of defense to break the overall defense.

The first logic locking scheme is designed to mitigate the threat of SAT attack; details of the inner workings of SFLL-rem are available in~\cite{sfll-rem}. 
SFLL-rem uses point-functions to increase the effort of the SAT attack by eliminating one key in each iteration out of a possible 2\textsuperscript{N} keys, where N is the number of bits in the key.
Although it is SAT attack resistant, this defense is vulnerable to approximate attacks. 

To thwart approximate attacks~\cite{appsat_host_2017,shen2017double}, the benchmarks implement RLL~\cite{RLL} on top of SFLL-rem to increase the output corruption when an incorrect key is used (i.e., to produce incorrect functionality for a partially recovered (approximate) key).  
RLL randomly inserts key gates into the netlist which may or may not at times provide 50\% output corruption. 
More sophisticated techniques such as fault-analysis based logic locking (FLL)~\cite{fll} and strong logic locking (SLL)~\cite{sensitization}  can be used to complement SFLL. 

\subsubsection{Combinational Locking Benchmark Preparation}
{\it Team X} applied SFLL-rem + RLL to 7 different circuits comprising two variants per reference design (i.e., one provided with an oracle, the other without) derived from academic benchmark circuits (taken from ITC~'99 benchmark suite), and a locked Cortex-M0 microprocessor (as a bonus challenge).  The circuits are named small, medium, and large, based on relative gate counts and listed in~\autoref{tab:circuits}. The team chose key sizes for each of the small, medium, large, and bonus circuits, according to their gate count. 
For the small, medium, and large circuits, we choose 40, 60, and 80 bits of security, respectively, in part to make the competition approachable. The bonus circuit, however, is locked with 128-bit security.

\begin{table}[t!]
\caption{Combinational circuit benchmarks\label{tab:circuits}. Small benchmarks are locked with a 40-bit RLL key and 40-bit SFLL-rem key. Medium benchmarks are locked with a 60-bit RLL key and 60-bit SFLL-rem key. Large benchmarks are locked with 80-bit RLL key and 80-bit SFLL-rem key. The bonus circuit was locked with 128-bit RLL key and 128-bit SFLL-rem key. }
\centering

\resizebox{\columnwidth}{!}{%
\begin{tabular}{@{}C{4cm}ccc@{}}
\toprule
Competition benchmark & \# Inputs & \# Outputs & \# Gates \\ \midrule
small (derived from b20\_C) &  522 & 512 & 20226 \\
medium (derived from b22\_C) &  767 & 757 & 29951 \\
large (derived from b17\_C)&  1452 & 1445 & 32326 \\ 
bonus (Cortex-M0)& 892 & 1746 & 16164\\ \bottomrule
\end{tabular}%
}
\aftertable
\end{table}

The combinational locking process involves three steps:
\begin{enumerate}[leftmargin=*,noitemsep]
\item Remove all traces of the benchmark identity. The modifications prevent recovery of the functionality by comparing the circuit against public benchmarks\footnote{Cadence LEC can provide a patch file to recover original functionality.}.
\begin{enumerate}
\item Change gate types of randomly selected gates. 
\item Rename internal nets by converting circuit to AND-Inverter gate (AIG) format.
\item Rename I/O ports.
\end{enumerate}
\item Lock modified circuit with SFLL-rem.
\item Lock SFLL-rem locked circuit with RLL.
\end{enumerate}

{\it Team X} verified the unlocking of the design using an open source equivalence checker~\cite{sat-attack}; it requires the locked netlist, the original netlist, and the key value. 
To prepare an oracle for the locked circuits, {\it Team X} converted the modified benchfile to executable permitting access to I/Os and not the internal circuitry. This mimics access to a working chip obtained from the market.
Since the competition does not provide the original netlist, the red teams cannot use equivalence checks to verify their key bits. Hence, the red teams report back to the blue team (via the organizers) who confirm how many of the key bits recovered are correct.

\begin{table}[tb]
\centering
\caption{Statistics of the sequential circuit benchmarks
\label{tab:seq_circuits}}
\resizebox{\columnwidth}{!}{%
\begin{tabular}{@{}C{1cm}C{1.5cm}C{1.5cm}C{1.5cm}C{1.5cm}C{1.5cm}@{}}
\toprule
Attack Model & Competition Benchmark & Academic Benchmark & No. of Inputs & No. of Outputs & Length of \newline Key Sequence \\ \midrule
Warm-up & Tiny & Custom & 6 & 71 & 18 \\ \midrule
\multirow{3}{*}{\shortstack{With \\ Oracle}} & Small & i2c & 18 & 14 & 51 \\
 & Medium & md5 & 41 & 35 & 63 \\
 & Large & s35932 & 36 & 32 & 134 \\ \midrule
\multirow{3}{*}{\shortstack{Without \\ Oracle}} & Small & s15850 & 13 & 87 & 52 \\ 
 & Medium & s13207 & 11 & 121 & 103 \\
 & Large & uart & 21 & 13 & 162 \\ \bottomrule
\end{tabular}%
}
\aftertable
\end{table}

\begin{table*}[t]
\centering
\caption{Combinational Locking Attack Results.} 
\label{tab:comb-csaw-results}
\resizebox{\textwidth}{!}{%
\begin{tabular}{ccccccccccc}
\hline
& & & \multicolumn{8}{c}{Benchmark} \\ \cmidrule(lr){4-11}
\multirow{2}{*}{Team} & \multirow{2}{*}{Approach} & \multirow{2}{*}{Attack Scenario}  & \multicolumn{2}{c}{Small (40+40)} & \multicolumn{2}{c}{Medium (60+60)} & \multicolumn{2}{c}{Large (80+80)} & \multicolumn{2}{c}{Bonus (128+128)} \\
 & & & RLL & SFLL & RLL & SFLL & RLL & SFLL & RLL & SFLL  \\ \hline
A & Key Sensitization & Oracle & 40/40 & - & 60/60 & - & 80/80 & - & - & -  \\
B & Hamming Distance-based Attack & Oracle & 30/30 & - & 50/50 & - & 72/72 & - & - &  - \\
C & Automated Analysis + SAT & Oracle & 11/18 & - & 31/50 & - & 10/34 & - & - & -\\
D & Sub-circuit SAT & Oracle& 17/17 & - & 29/29 & - & - & - & - & -  \\
E & Redundancy-based & Oracle-less & 28/28 & 4/12 & 35/35 & 23/28 & 45/45 & 0/51 & 66/66 & 8/27  \\
F &  Bit-flipping Attack & Oracle & 40/40 & - & 60/60 & - & 80/80 & - & - & -  \\
G & Topology guided attack & Oracle-less & 15/32 & - & 19/50 & - & 36/73 & - & 75/108 & -  \\ \hline
\end{tabular}%

} 
\end{table*}

\begin{table*}[t]
\centering
\caption{Sequential Locking Attack Results.} 
\label{tab:seq-csaw-result}
\begin{tabular}{C{1cm}C{2cm}C{2cm}C{2cm}C{2cm}C{2cm}C{2cm}C{2cm}}
\hline
& \multicolumn{2}{c}{Small} & \multicolumn{2}{c}{Medium} & \multicolumn{2}{c}{Large} & \\ \cmidrule(lr){2-3} \cmidrule(lr){4-5} \cmidrule(lr){6-7}
Team & \# input ports used for key assertion & Length of key sequence (\# clock cycles) & \# input ports used for key assertion & Length of key sequence (\# clock cycles) & \# input ports used for key assertion & Length of key sequence (\# clock cycles) & Attack Scenario \\ \hline
\multirow{2}{*}{A} & 5/10 & 10/52 & 4/5 & 12/103 & 4/19 & 11/162 & Oracle-free \\
 & -/5 & -/63 & 5/16 & 10/51 & 4/26 & 10/134 & Oracle \\ \hline
\end{tabular}%
\aftertable
\end{table*}

\subsubsection{Sequential Locking Overview}
{\it Team Y} uses an improved 
variant of  state space obfuscation \cite{chakraborty2009harpoon} for sequential locking. 
Since the correct operation of any IP depends on the control logic (i.e., a function of inputs and control states embedded in a finite state machine (FSM) into the design) the sequential locking defense involves expanding the functional state space by inserting additional state registers into the FSM. The added and the existing state registers form a combined FSM which dictates the  \textit{normal} or \textit{locked} operating mode of the design.

\begin{figure}[tb]
\centering
\subfigure[]{\includegraphics[width=0.42\columnwidth, trim = 10 30 10 30, clip]{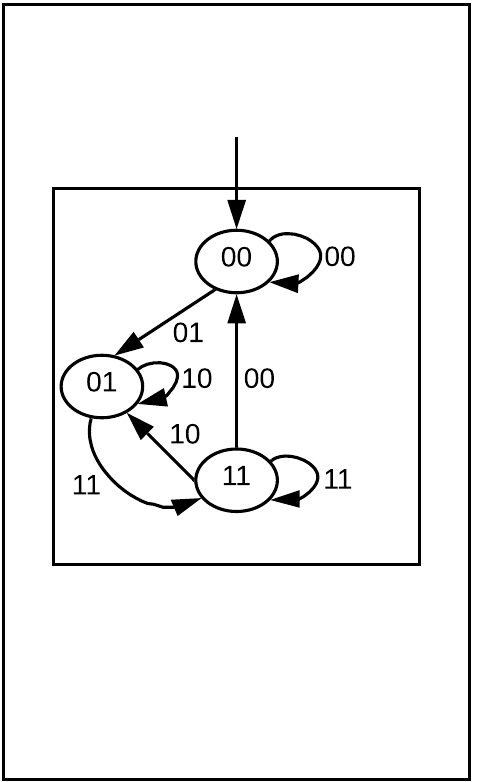}\label{fig:seq-orig}}
\subfigure[]{\includegraphics[width=0.49\columnwidth]{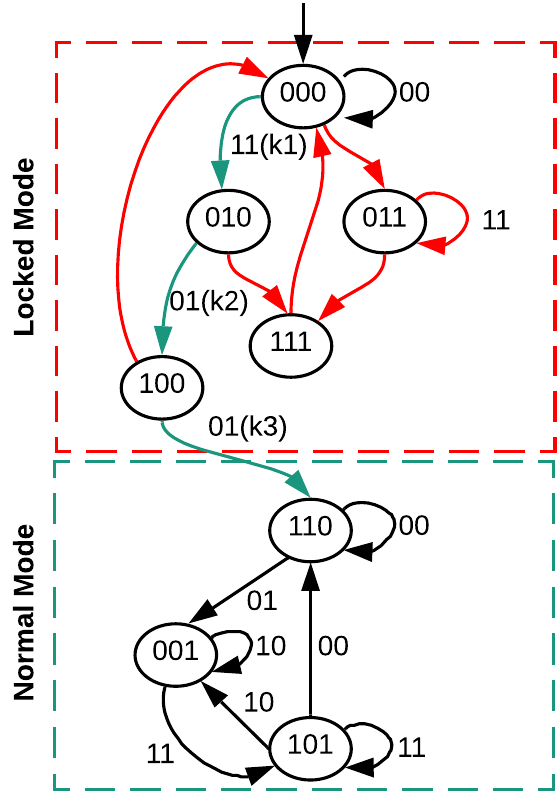}\label{fig:seq-locked}}
\caption{(a) Original state machine (b) Locked state machine with added state register (increasing number of reachable states from three to eight)\label{fig:seq_obf}}
\afterfig
\end{figure}

To enable the normal mode of operation, an authorized user needs to apply a sequence of unlocking key inputs through the design's primary inputs.  This causes the extended state machine to traverse along a specific "unlocking" state sequence. 
Without the correct key sequence, the locked design will remain in, and transition between, a set of non-functional states, producing non-functional (i.e., corrupted) outputs. 
Thus, to enable the normal operation mode, an attacker would:  
\begin{enumerate}[leftmargin=*,noitemsep]
    \item Find the inserted state registers used in locking,
    \item Find the unlocking key sequence, and
    \item Extract the functional states from the locked design.
\end{enumerate}
The locking mechanism is illustrated in~\autoref{fig:seq_obf}. 
To enable the normal operation, the \textit{green path} must be traversed by applying a sequence of key inputs \textit{K1, K2, and K3} where each of the key inputs can comprise varying subsets of the primary inputs. If an incorrect key input is applied, the \textit{red path} is traversed and the normal mode is never reached. 
As the IP is in the locked mode, it remains non-functional. Unlike combinational locking, sequential locking re-uses primary inputs as key inputs. Based on the constraints and level of the security, the key width and the length of the key sequence is altered. 

\subsubsection{Sequential Locking Benchmark Preparation}
Each design is locked using different lengths of key sequence as shown in~\autoref{tab:seq_circuits}.
To lock a design, extra state registers are inserted to increase the reachable state space of the design. 
The expanded state space overlaps with the existing state space; the state transition function of the existing state machine is modified to use the additional states available from the extra state registers, including the addition of the unlocking input sequence. 
The number of gates, the number of inputs/outputs and the number of state registers in the design determine the complexity of  locking and difficulty of unlocking. 

The locked designs are synthesized using LEDA 250nm technology node. {\it Team Y} renamed the wires and cells to prevent attackers from isolating the obfuscation logic by looking at the names. 
However, primary input/output names were kept as is.  The attackers are given white-box access to the locked netlists, \emph{i.e.}, they have access to the inputs, outputs, and individual cells of the design. For oracle-less attacks, the attackers were given the locked netlist.  For oracle-based attacks, a locked netlist and the original netlist (i.e., the design before locking) were provided.  To verify the results, the attacker can simulate the locked design by applying the key sequence, followed by a random pattern. The output is compared to the output of the oracle for the same input. 

\section{Benchmarking Results}
\label{sec:csaw}
    
\subsection{Results Overview}
    We present the finalists' attack results on combinational and sequential locking in~\autoref{tab:comb-csaw-results} and~\autoref{tab:seq-csaw-result}, respectively.  
    In the combinational locking attacks, several teams recovered all RLL key bits. Each table presents the teams, their attack approach and their chosen attack scenario (oracle or oracle-less). This is followed by attack success on the various benchmarks represented by the number of key bits recovered. Red teams reported that their attacks recovered a number of key bits. The blue team checked to see how many matched up. 
    We present the result as "x/y", i.e., the number of bits, x, verified as correct, out of the bits recovered by the attacker, y.  No team was able to recover the complete set of SFLL keys. 
    
    In the attacks on sequential locking, none of the teams reported an entirely correct sequence and set of key input ports. One attack recovered a subset of the key input ports used for key assertion---in~\autoref{tab:seq-csaw-result}, this is represented as "a/b", where a out of the b input ports used as key inputs were correctly identified. 
    The attack attempted to identify the key sequence; this is reported as "c/d" where c is the sequence length recovered by the attack, while d is the actual length. 

\subsection{Outline of Attack Approaches}
The red teams proposed interesting attack directions, in some cases incorporating those proposed in the literature. 
None of the attacks were able to successfully recover the entire key. 
In this section, we outline the attack attempts and their success.  The appendix presents more detail on the attacks. 

\begin{itemize}[leftmargin=*,noitemsep]
\item \textbf{ATPG-based} (Oracle-guided)
\textit{Team A} applied the sensitization attack~\cite{sensitization} on combinational locking. 
While the team identified the primary inputs connected to the key inputs, they failed to recover the SFLL-rem key.
\textit{Overall, they recovered 100\% of the RLL key bits and none for SFLL-rem.} The team also used an ATPG-based method to attack sequential locking based upon~\cite{duvalsaint_characterization_2019}, but this did not correctly recover the key sequence. 

\item \textbf{Hamming Distance (HD)-based Attack} (Oracle-guided) 
\textit{Team B} proposed a divide-and-conquer approach to attack the combinational locking. 
To attack SFLL-rem, they pick the protected input patterns (PIP) one at a time and apply the SAT attack. 
\textit{This team was able only to recover most of the RLL key bits and none of the SFLL-rem key bits.}

\item \textbf{Automated SAT} (Oracle-guided)
\textit{Team C} attempted a divide-and-conquer approach by dividing the circuit into smaller logic cones, starting with primary outputs and performing a SAT attack on the smaller cones, focusing on  cones with fewer key inputs. 
\textit{The team recovered fewer than 50\% of RLL key bits and no SFLL-rem key bits.} 

\begin{table*}[tb]
\centering
\caption{Important findings (SAT: Satisfiability solver; ATPG: Automatic Test Pattern Generation).}
\label{tab:my-table}
\resizebox{\textwidth}{!}{%
\begin{tabular}{@{}L{2.4cm}L{2.9cm}L{2.9cm}L{2.9cm}L{5cm}@{}}
\toprule
Locking Technique & On SAT-based Attacks & On ATPG-based Attacks & On Structural Analysis & Important Takeaways\\ \midrule
SFLL-rem + RLL (Combinational) & Remains the technique-of-choice for attacking LL. Successful in attacking RLL.   & Emerging attack formulation, showing some promise even without oracle access & Crucial for attacking locked designs in a more scalable, piece-wise manner & Most attempted attacks require oracle access; the potential for oracle-free approaches require further exploration. RLL can be attacked successfully while SFLL-rem remains resilient to attack \\
\hline
State Space Obfuscation (Sequential) & Not attempted, requires scan access or appropriately unrolled design & Also requires unrolling, but \# of cycles to unroll is exponential in number of flip-flops and initial (reset) state & Important to find the added state registers & Not yet as mature/familiar as combinational locking, perhaps requiring further dissemination for a more thorough evaluation. \\ \bottomrule 
\end{tabular}%
}
\aftertable
\end{table*}

\item \textbf{Sub-circuit SAT} (Oracle-guided)
\textit{Team D} proposed a  SAT attack on combinational locking by simplifying the circuit using two strategies and then applying SAT attack. First, they find an output which depends on one key input. Second, they find a key input which propagates to one output only.  
\textit{They recovered fewer than 50\% of the correct RLL keys and none of the SFLL-rem key bits.}

\item \textbf{Circuit Redundancy} (Oracle-less)
{\it Team E} centered their attack on circuit redundancy~\cite{redundancy_attack} by observing that a post-synthesis insertion of key gates yields invalid design alternatives that do not adhere to well-established design characteristics, specifically the redundancy level, thus giving away the incorrect key bit values.
\textit{The team recovered more than 50\% of RLL key bits, all correctly.  
They recovered fewer than 50\% of the SFLL-rem key bits.}

\item \textbf{Bit-flipping attack} (Oracle-guided)
\textit{Team F} devised an attack on the combinational locking using bit flipping~\cite{bitflip}. 
\textit{ This attack recovered all RLL key bits in under a minute for the largest benchmark and none of the SFLL-rem key bits.}

\item \textbf{Unit Function Search Attack} (Oracle-less) %
{\it Team G} attempted the Unit Function search attack on the combinational locking ~\cite{zhang2019tga}. 
If one or more key gates are placed in an instance of repeated unit function (\textit{UF}) during the locking of a circuit, the original netlist can be recovered by searching the equivalent unit functions (\textit{EUFs}) with all hypothesis keys. 
\textit{This attack recovered fewer than 50\% of RLL key bits, and no SFLL-rem key bits.} 

\end{itemize}

\section{Reflections and Lessons Learned}
\label{sec:discuss}

Following the benchmarking effort, we reflected on the undertaking and solicited feedback from participants and judges.  This section describes observations and insights we drew from examining the competition submissions and continuing discussions amongst the coordinators, red-team and blue-team participants, judges, and external observers.

\subsection{Observations and lessons for logic locking}

\textbf{Important Findings for Logic Locking.} %
    \autoref{tab:my-table} summarizes the findings from the benchmarking process. 
    SAT attacks remain a popular element of the attacks on SFLL-rem + RLL, but requires oracle access. Exploration on oracle-free approaches is an open challenge.  State space obfuscation presents interesting challenges for attackers, especially without access to a scan-chain; attackers need to decide how much to unroll a design to perform analysis. Sequential locking is less familiar to the community and hence is less mature and will benefit from wider dissemination. 
    As it stands, SFLL-rem and State Obfuscation remain "unbroken" based on this benchmarking effort, with the fielded attacks offering only partial "success" in key recovery. 

\textbf{On Participant Prior Experience.} %
    Most participants in the benchmarking exercise were experienced researchers in VLSI testing, hardware security, and specifically logic locking. 
    Several teams featured members with a mix of background familiarity, where undergraduates worked with doctoral candidates that had published works on attacks/defenses in logic locking. 
    In the qualifying round, some teams comprised students who were introduced to the concepts of logic locking in class a few weeks before the contest. 
    Furthermore, most teams concerned themselves with combinational locking only---suggesting that sequential locking is less familiar. 
    
\textbf{On Tools Used.} %
    Participants described a number of different tools used to perform attacks on the benchmark circuits. 
    These include various SAT solvers, ATALANTA ATPG tool~\cite{atalanta}, and a lot of scripting (typically in languages like Python or Perl). 
    Many teams automated netlist parsing and analysis, although some teams started with a level of manual inspection.

\subsection{Reflections on the benchmarking process}
\textbf{On Logistics and Communication.} %
    Throughout the benchmarking process, the coordinators acted as intermediaries between red teams and blue teams, and were keeping all people in the loop. The primary medium for communication was e-mail; participants were invited to maintain a live report to detail their efforts throughout the exercise, although these were often left untouched until the hours before the deadline. Ensuring that e-mails were received and noted is particularly challenging, although perhaps understandable given that participants may have had numerous competing concerns (i.e., as graduate students, many participants had coursework and other academic pursuits with respect to the academic calendar). Maintaining interest and incentivizing participation are difficult challenges; we tackled this by being in frequent contact with the red teams, blue teams and the judges\footnote{We offered a funded trip to the in-person finals for the student finalists as additional motivation.}. 

\textbf{On Benchmark Management.} %
    There were a number of challenges in terms of benchmark preparation, including how best to prepare and validate the benchmark circuits, and subsequently, how to distribute them. 
    In our benchmarking effort, we relied on good-faith, best-effort practices from the blue teams. 
    The blue teams independently prepared benchmarks and accompanying support material (e.g., oracles) after consultations with the community. 
    Once the benchmarking commenced, the coordinator facilitated communication between red teams and blue teams, accommodating requests where reasonable (for example, recompilation of oracles for different execution platforms).     While this approach worked well, future iterations will benefit by additional documentation and vetting of the benchmark circuit preparation process by parties external to the blue teams.  
    Furthermore, future iterations could adopt standard formats for benchmarks (e.g., netlists in Verilog or VHDL) synthesized with an open-source technology library, for better tool interoperability. 

\textbf{On Result Assessment/Verification.} %
    Red teams submitted their attack outputs to the coordinator; these were then passed on to the blue teams for correctness assessment. 
    While this process works in terms of assessing key recovery success, our benchmarking effort was not able to evaluate other metrics; while the scalability of an attack can be somewhat inferred by considering which of the benchmarks were attacked, other dimensions like attack implementation runtime or space complexity were not measurable (i.e., we did not offer a common computation platform). 
    In some ways, we can interpret the results of this effort as a rough measure of what an attacker might be able to achieve end-to-end within $\sim$3 months, albeit with several caveats (e.g., teams were not working exclusively on the problem during the competition). 

\textbf{On Resources.} %
    The predilection towards combinational locking suggests that there is less familiarity with sequential locking in the community (at least, as represented by the participants). 
    This raises a bigger question about what sort of resources and documentation could be provided to participants in future iterations to reduce any barriers to entry (and hopefully, encouraging more robust evaluation of the locking techniques). 
    It would also be interesting to investigate what additional collateral material might support further evaluations of logic locking in "real-world" scenarios.  For example, datasheets or other design documents as "plausible" supporting information that is accessible to an attacker; access to such information might enable a thorough examination of logic locking. 

\subsection{Perspectives on Application-level Directions}
    While the red/blue team exercise was underway, we supported an orthogonal exploration by \textit{Team Z}. 
    This team proposed an alternative attacker model that provides a promising new direction for logic locking research. Their approach aimed to quantify application-level ramifications of logic locking, considering each locked benchmark netlist as a small part of a larger and more complex system with a specific  application.
    Because logic locking impacts the IC as a whole, application-level considerations provide important context for both combinational and sequential locking. With this context, architectural factors, such as error tolerance, and application-level factors, such as workload characteristics, must now be considered. This presents a holistic view of logic locking in practice.
    
    To explore this attacker model, \textit{Team Z} proposed a custom logic locking simulation framework, ObfusGEM. ObfusGEM integrates logic locked netlists within a cycle-accurate GEM5  model of a processor~\cite{binkert2011gem5}. By running arbitrary workloads on this model and tracking the divergence between a locked and unlocked processor, the application-level impact of logic locking is quantified \cite{performancelocking}. This provides a promising  avenue to evaluate logic locking moving forward.

    During this benchmarking, \textit{Team Z} used ObfusGEM to integrate custom, small netlists, locked with SFLL-rem, into an x86 processor IC. Using this model, they showed that approximate keys, sufficient for the correct execution of a variety of software benchmarks, could be located, despite the presence of incorrectly keyed logic locking. While the locked netlists were of a smaller scale compared to the main benchmarking exercise, these results point at application-level context as a potential new direction for both attack and defense research, including hardware/software co-design and high-level synthesis, that can leverage application-level considerations to strengthen the security of logic locking.  Thus, in future evaluations, the community should examine application-level considerations and the context they provide for combinational and sequential logic locking. 
    
    Several teams were able to recover  the RLL keys for the combinational locking benchmarks, arguing that the remaining key inputs could be set randomly to retrieve an approximate circuit, with low error rate.  This points to a well-known aspect of SFLL-based locking, where output corruption is limited to combat oracle-guided attacks.  However, implications of an approximately recovered circuit is application dependent. Since different input/output pairs may have different levels of "importance" depending on the context, generalizing that the recovery of a circuit with a low error rate constitutes a successful attack may not be appropriate. How best to reflect application-level concerns in the next iteration is a direction to pursue. 
    
    \subsection{Broader Reflections for Hardware Security}
    As mentioned in~\autoref{sec:intro}, logic locking is one of several design-for-trust techniques. 
    The lessons from our critical examination of the literature and benchmarking approach provide insights into the challenges faced in hardware security generally, and may in turn inform future efforts in evaluating alternative, emerging techniques in this area. 
    
    As we discussed in~\autoref{sec:preliminaries}, selecting appropriate benchmarks is crucial for characterizing techniques (e.g., with respect to overheads) and providing context for how effective an attack/defense is. In hardware security, more broadly, procuring benchmarks is not trivial. 
    Unlike software, where there is ample (and production-quality) open-source resources, the hardware design community is not as populous (for example, there are $\geq$200K active Java projects on GitHub~\cite{git-hub-stats}, whereas OpenCores contains $\sim$1K hardware IP-blocks~\cite{open-cores-stats}). 
    
    Having access to open-source tools might help with benchmark creation and evaluation. 
    However, while open-source academic tools for hardware design exist (e.g., ABC~\cite{abc} for logic synthesis), industrial electronic design automation (EDA) tools are less accessible, and this may preclude fully open-source logic locking tool-chains. Having said that, efforts such as OpenROAD~\cite{ajayi_toward_2019}, whose goal is an open-source RTL-to-GDSII flow, may change the future prospects. 
    
    Another area for future efforts include greater industry involvement, notwithstanding possibly complex legal IP issues or non-disclosure agreements, etc. 
    For instance, the Hack@DAC~\cite{dessouky_hardfails:_2019} organizers were able to find a way for industry engineers to contribute meaningful hardware bugs to an open-sourced SoC design. 
    While not security related, the long running ICCAD CAD contests~\cite{iccad-cad-contest} have shown success in bringing industry and academia together, such as the industry-provided benchmarks for physical design in~\cite{nam_ispd2005_2005}. 
    How best to pursue this for design-for-trust techniques generally is an open question. Further, incorporating application-related context will be useful (e.g., identifying and quantifying important inputs and outputs). 
    Industry could propose different benchmarks for characterizing different facets of the techniques, perhaps locking for datapaths, locking for sea-of-gates, or locking for SRAMs. 
    
    Our work has demonstrated the value of coordinated evaluation of hardware security techniques. With industry, government, and academic support, logic locking and other hardware security techniques can benefit from formal and ongoing evaluation. By making these processes regular and structured, researchers could submit new techniques on an ongoing basis for rigorous assessment. Such a process would increase confidence in hardware security technologies. 

\section{Future Outlook}
\label{sec:future}
    
    \textbf{Next Steps for Benchmarking.} %
    There are a number of areas that future iterations of community benchmarking can include for greater insights into the usability, practicality, and resilience of logic locking techniques. 
    These include: (1)~a common (possibly cloud-based) platform for comparing attack techniques: having a standard computing platform for launching attacks will allow better side-by-side comparisons of attack strategies with respect to their execution times, scalability, etc., (2)~an even wider variety of benchmark circuits, possibly grouped by application domains (with this additional context provided to or withheld from the red teams depending on the threat model to be emulated), and (3)~varying the amount of information provided to the attacker (e.g., application context). 
    In our benchmarking exercise, we treated the combinational and sequential locking techniques in isolation---in the next iteration, it could be worth expanding the fielded defenses as well as investigate if we can gain more from combining techniques. The SFLL-rem technique was combined with RLL and the RLL-keys were broken. Perhaps other locking approaches can be mixed-in successfully. 
    
    \textbf{Open-sourcing and Support Materials.} %
    For deeper scrutiny and wider adoption of logic locking, the natural next step will be fully open-source implementations of the locking techniques with detailed support materials (algorithm descriptions, user guides, etc.). This may pave the way for standardization, modeled after the NIST-style processes where the community-at-large publicly dissects the algorithm \textit{and} its (reference) implementation. 
    Increasing the visibility into the inner workings of locking techniques will increase confidence in correctness, and move the needle even closer towards Kerckhoff's principle in favor of security by obscurity. 
    This is a challenge as locking techniques may use commercial tools as part of the tool flow.  License costs may be prohibitive to some and non-disclosure agreements may hold back others.
    
    \textbf{Open Questions.} %
    While this paper explored shortcomings of the evaluation in the logic locking literature and attempted to address these through our foray in benchmarking, more open questions remain. 
    These include ongoing questions about what constitutes a successful attack, i.e., how can we formalize meaningful notions of approximate recovery or application-level concerns? 
    More work could also look to separating analysis of a locking technique from attacks that arise from a flawed implementation of the technique. 
    Furthermore, is it possible to definitively conclude that a defensive technique has "graduated" from this process? 
    This will inform the steps that could be taken after a technique succeeds at this process, in terms of potential impacts on wider adoption or public policy. 
    
    \textbf{Concluding Remarks.} %
    In this paper, we prepared, ran, and reflected on the first benchmarking effort in logic locking. 
    Through this process we worked towards leveling the playing field where defenders and attackers were given the opportunity to "put their best foot forward". 
    As it stands, SFLL-rem and State-based locking remain "unbroken" based on this benchmarking effort, with the fielded attacks offering only partial success. 
    Our efforts produced a timely snapshot of the current state-of-the-art in logic locking for digital design IP protection. 
    As these techniques mature and new attacks and defenses emerge, the lessons learned from this community effort will provide a foundation for future endeavors. 
   
    \section*{Acknowledgments}
    B. Tan is supported in part by ONR Award \# N00014-18-1-2058. R. Karri is supported in part by National Science Foundation (NSF) Grant \#1513130, \# N00014-18-1-2058 and NYU/NYUAD Center for Cyber Security.   
    O. Sinanoglu, N. Limaye and A. Sengupta are supported in part by the Defense Advanced Research Projects Agency (DARPA) under Grant HR001116S0001-FP39, and in part by the NYUAD Center for Cyber Security.
    S. Bhunia and M. Rahman were supported in part by the DARPA OMG Program grant.
    Z. Han, A. Benedetti, L. Brignone, M. Yasin, and J. Rajendran are supported by the Office of Naval Research (ONR Award \#N00014-18-1-2058), the National Science Foundation (NSF CNS-1822848), and the Defense Advanced Research Projects Agency (DARPA HR001116S0001-FP39). 
    This research was developed with funding from the Defense Advanced Research Projects Agency (DARPA)

    \textbf{Disclaimer.} The views, opinions and/or findings expressed are those of the author and should not be interpreted as representing the official views or policies of the Department of Defense or the U.S. Government.

\IEEEtriggercmd{\balance}
\bibliographystyle{IEEEtran}
\bibliography{IEEEabrv,x_references}

\appendix

\section{Details of Attack Approaches\label{sec:attack-detail}}
While some red teams proposed new attacks, others used attacks published in the literature. None of the attacks successfully recovered the entire key. This section provides further details on the attack attempts (in no specific order). 

\textbf{ATPG-based} (Oracle-guided)
{\it Team A} applied the sensitization attack~\cite{sensitization} on the combinational locking benchmarks. In this attack, automatic test pattern generator (ATPG)\footnote{ATPG is used to test circuits to detect manufacture-time faults such as a wire stuck-at-zero and stuck-at-one.} is used to detect a fault at a key input, with the other keys tied to a don't-care. If a test pattern is found, this indicates that there is a path from the key input to a primary output that does not require setting the other keys.  The test pattern is applied to the oracle to propagate the value of the key input to the primary output.  To recover the SFLL-rem key bits, a different approach based on fault equivalence and dominance was attempted. 
This attack looks for a fault that is equivalent or dominant to the fault that was removed. While the team identified the primary inputs connected to the key inputs, they failed to recover the SFLL-rem key.
\textit{They recover 100\% of RLL key bits and none of the SFLL-rem key bits. }

The team used an ATPG-based method to attack the sequential locking benchmarks, based upon~\cite{duvalsaint_characterization_2019}.  
To attack a locked sequential circuit, the ATPG-based attack unrolled the circuit (as scan-chain access was not provided). 
Unrolling a sequential circuit increases the attack complexity exponentially and this complexity increases with the number of registers.  In order to deploy a successful attack, the circuit needs to be fully unrolled.  The team incorrectly guessed the number of cycles to unroll to be 5 cycles. This led to an incomplete unrolling and a failure to recover the complete key. 

\textbf{Hamming Distance (HD)-based Attack} (Oracle-guided) 
{\it Team B} proposed a divide-and-conquer approach to attack the combinational locking. There are three steps to perform this attack: identify the type of key inputs, strip partial RLL key inputs, and launch the attack.
\begin{enumerate}[leftmargin=*,noitemsep]
\item  Split the locked netlist into individual logic cones (ILCs). 
\item Count the number of key bits in each cone to identify the type of protection for each ILC and the type of each key input. Since they find all ILCs and each cone's protection type, they can find the RLL key value. 
\item Select all RLL cones and merge them into a netlist with fewer primary outputs (compared to the full netlist).  The SAT attack on this netlist returns a valid RLL key with partial RLL key inputs.
Applying this key to the original locked netlist produces a simplified locked circuit. 
\end{enumerate}

To attack the SFLL-rem lock, they pick one cone which is only protected by SFLL, as shown in~\autoref{alg:hd_based_attack}. First, they extract the functionality stripped circuit (FSC) from the locked cone. 
From this they collect a set of candidate protected input patterns (PIPs) whose Hamming distances are no greater than a  threshold $d$ to at least one PI in the FSC's reduced PI table. 
The PIP candidates are fed into the oracle and the FSC to identify differences in the output, thus verifying that the input pattern is indeed protected. 
If they can find one verified PIP, they use this PIP as the first input pattern into the SAT-based attack and find the correct key from the attack. This key is valid for other locked cones.
\textit{This team recovered most of the RLL key bits and none of the SFLL-rem key bits.} 

\begin{algorithm}
\begin{algorithmic}[1]
\caption{Hamming Distance (HD)-based Attack}
\label{alg:hd_based_attack}
\REQUIRE A SFLL-fault cone $\mathcal{C}_{locked}$, oracle $\mathcal{O}$, parameter $d$
\ENSURE Correct key $key_c$
\STATE $\mathcal{C}_{FSC} \gets \text{extract\_FSC} (\mathcal{C}_{locked})$
\STATE $Reduced\_PIT \gets \text{extract\_PI\_table}(\mathcal{C}_{FSC})$
\STATE $PIP_{cand} \gets \{p | HD(p,pi)\le d, \exists pi \in Reduced\_PIT\}$
\FOR{$p \in PIP_{cand}$}
    \IF{$\mathcal{O}(p) \neq \mathcal{C}_{FSC}(p) $}
        \STATE $key_c\gets\text{SAT\_simulation}(\mathcal{C}_{locked},\mathcal{O},p)$
        \RETURN $key_c$
    \ENDIF
\ENDFOR
\end{algorithmic}
\end{algorithm}

\textbf{Automated SAT Attack} (Oracle-guided)
{\it Team C} attempted a divide-and-conquer approach by dividing the circuit into smaller logic cones, starting with primary outputs and identifying the fan-in logic. 
The team automated this structural-level analysis and performed a SAT attack on the logic cones, focusing on those cones with fewer key inputs. The circuit is divided into individual logic cones. An analysis of the cones reveals that some outputs depend on only one key input and a relatively small number of inputs. As an intermediate step, they attempted to find DIPs on a small logic cone with one key input, 33 primary inputs, and one primary output. The SAT solver was unable to solve the reduced circuit, so a key sensitization attack on the smaller logic cone was performed. This attack chooses an arbitrary set of inputs, executes the oracle, collects the targeted output, and  runs the SAT solver with the key input as the unknown. The SAT solver produced an output for the single unknown value. After finding some keys from logic cones with a single key input, the team targeted logic cones with increasing numbers of key inputs and one unknown key input.
\textit{The team recovered $<$50\% of RLL key bits and no SFLL-rem key bits.} 

\textbf{Sub-circuit SAT Attack} (Oracle-guided)
{\it Team D} launched a SAT attack on combinational locking. They used two strategies to find a sub-circuit on which to apply SAT attack. \begin{enumerate}[leftmargin=*,noitemsep]
\item 
They find an output which depends on one key input. Then, the SAT attack is applied for the sub-circuit. If the SAT solver iterates twice to find the value of a key, it indicates that this value is correct. 
\item They uncover a key input which propagates to only one output. A SAT attack is mounted on the sub-circuit involving the cone of influence of this output.  The resulting key is correct as it depends only on that output and this value can be used to find additional key inputs recursively.
\end{enumerate}
\textit{They recovered $<$50\% of the RLL key bits and none of the SFLL-rem key bits.}

\textbf{Redundancy attack} (Oracle-less)
{\it Team E} used the redundancy attack~\cite{redundancy_attack} on the combinational locking benchmarks.
The attack is based on the observation that key gates modifies the netlist after synthesis and therefore produce arbitrary invalid design options that fail to adhere to certain design principles such as redundancy removal.
The attack dismantles the complexity of the key space by deciphering key bits individually or in pairs.
Since RLL key bits are inserted in the middle of the netlist, they have stronger local impacts on the redundancy level of nearby regions where a single modification to the intertwined re-convergent structures alone results in untestable faults.
On the other hand, incorrect key values for SFLL result in untestable faults at the convergence point of the functionality stripped circuit and recovery unit.
The removal of such redundant faults would make certain output bits completely unprotected by SFLL, thus proving the invalidity of the key assignment.
\textit{The team recovered $>$50\% of RLL key bits and $<$50\% of the SFLL-rem key bits.}

\balance

\textbf{Bit-Flipping Attack} (Oracle-guided)
{\it Team F} adopted the bit-flipping attack~\cite{bitflip} on the combinational locking benchmarks based on. 
The key bits of RLL and SFLL-rem are separated by fixing the key values to random values with the Hamming distance equal to one, and counting the number of Distinguishing Input Patterns (DIPs). 
DIPs are used to differentiate between the design outputs when different key values are applied. Since the error rate of SFLL-rem is exponentially low, the protected input patterns are rarely applied even if the SFLL-rem key is wrong. Thus, they randomly fix the SFLL-rem key and solve the RLL key by applying the SAT attack~\cite{sat-attack}. This attack recovers all the RLL key bits in under a minute for the largest benchmark. Algorithm \ref{alg:bit-flipping} shows the methodology. \textit{They recovered all RLL key bits and none of the SFLL-rem key bits.}

\begin{algorithm}
\begin{algorithmic}[1]
\caption{Bit-flipping Attack}
\label{alg:bit-flipping}
\REQUIRE Encrypted circuit C(X, K, Y) and oracle $eval.$
\ENSURE Encryption key $K_c. $
\STATE i = 1
\STATE $F_1 = C(X, K_1, Y_1) \wedge C(X, K_2, Y_2)$
\STATE \emph{Fixing SFLL-rem keys in} $K_1$ and $K_2$ \emph{to a random value}
\WHILE {$sat[F_i \land (Y_1 \neq Y_2)]$}
\STATE $X_i = sat\_assignment_{X}(F_i \wedge (Y_1 \neq Y_2))$
\STATE $Y_i = eval(X_i)$
\STATE $F_{i+1} = F_i \wedge C(X_i, K_1, Y_i) \wedge C(X_i, K_2, Y_i)$
\STATE $i = i + 1$
\ENDWHILE
\STATE $K_c = sat\_assignment_{K1}(F_i)$
\end{algorithmic}
\end{algorithm}

\textbf{Unit Function Search Attack} (Oracle-less attack) 
{\it Team G} mounted the Unit Function Search attack on the combinational locking benchmarks~\cite{zhang2019tga}.
If one or more key gates are placed in an instance of repeated unit function (\textit{UF}) during the locking of a circuit, the original netlist can be recovered by searching the equivalent unit functions (\textit{EUFs}) with all hypothesis keys. The hypothesis key bit will be the actual secret key bit if a match is found. The attack fails when the search fails to find a match with all hypothesis keys. 

This attack uses an efficient depth-first  search to find the EUFs in a locked netlist. Searching the EUFs in the netlist is equivalent to the subgraph isomorphism problem. Hence, they convert the netlist to a directed graph, where each gate in the netlist is a vertex, and each wire is an edge. For each EUF, the search algorithm traverses the generated graph to check for the existence of the same structure. 

Since each key bit is targeted individually, the average time to determine a secret key bit is in the order of seconds. 
The initial version of this attack in~\cite{zhang2019tga} targeted RLL presented in~\cite{roy_epic:_2008}, where there is no inter-dependency among key bits. 
\textit{This attack recovered $<$50\% of RLL key bits, and no SFLL-rem key bits.}

\end{document}